\documentstyle[aps,prl,epsfig]{revtex}

\draft

\begin{document}

\title{\bf Mirror quiescence and high-sensitivity position measurements
with feedback}

\author{
David Vitali$^1$, Stefano Mancini$^{1,2}$, 
Luciano Ribichini$^{3}$, and Paolo Tombesi$^1$}

\address{
$^1$INFM, Dipartimento di Matematica e Fisica,
Universit\`a di Camerino,
I-62032 Camerino, Italy
\\
$^2$INFM, Dipartimento di Fisica,
Universit\`a di Milano,
Via Celoria 16, I-20133 Milano, Italy
\\
$^{3}$Albert Einstein Institut f\"ur Gravitationsphysik, Aussenstelle
Hannover, Callinstrasse 38, D-30167 Hannover, Germany}

\date{Received: \today}

\maketitle
\widetext

\begin{abstract}
We present a detailed study of how phase-sensitive feedback schemes
can be used to improve the performance of optomechanical devices. 
Considering the case of a cavity mode coupled to an oscillating mirror 
by the radiation pressure, we show how feedback can be used to reduce 
the position noise spectrum of the mirror, cool it to its quantum ground 
state, or achieve position squeezing.
Then, we show that even though feedback is 
not able to improve the sensitivity of stationary position
spectral measurements, it is possible to design a 
nonstationary strategy able to increase this sensitivity.
\end{abstract}


\section{Introduction}

Mirrors play a crucial role in a variety of precision measurements
like gravitational wave detection \cite{GRAV} and atomic force 
microscopes \cite{AFM}.
In these applications one needs a very high resolution for position 
measurements and a good control of the various noise sources, 
because one has
to detect the effect of a very weak force \cite{caves,onofrio}. 
As shown by the pioneering work
of Braginsky \cite{BRAG}, even though all classical noise sources
had been minimized, the detection of gravitational waves 
would be ultimately
determined by quantum fluctuations and the Heisenberg 
uncertainty principle.
Quantum noise in interferometers has two fundamental sources, 
the photon shot noise 
of the laser beam, prevailing at low laser intensity, 
and the fluctuations
of the mirror position due to radiation pressure, which is 
proportional to the incident laser power. This radiation 
pressure noise is the
so-called ``back-action noise'' arising from the fact 
that intensity fluctuations
affect the momentum fluctuations of the mirror, which are 
then fed back into
the position by the dynamics of the mirror. 
The two quantum noises are
minimized at an optimal, intermediate, laser power, 
yielding the so-called
{\em standard quantum limit} (SQL) \cite{caves,BK}.
Real devices constructed up to now are 
still far from the standard
quantum limit because quantum noise is much smaller 
than that of classical
origin, which is essentially given by thermal noise. 
In fact, present
interferometric gravitational wave detectors are 
limited by the Brownian
motion of the suspended mirrors \cite{ABRA}, 
which can be decomposed into 
suspension and internal (i.e. of internal acoustic modes) 
thermal noise.
Therefore it is very important to establish the
experimental limitations determined by
the thermal noise, and recent 
experiments \cite{HADJAR,TITTO}
go in this direction. 

Recently it has been reported \cite{SCI} the first experimental 
evidence 
of the reduction of thermal noise by means of the radiation pressure 
of an appropriately modulated laser light incident on the back of 
the mirror \cite{HEIPRL}.
The method was based on a phase-sensitive feedback control proposed 
in Ref.~\cite{MVTPRL}:
detect the mirror displacement through a homodyne measurement, 
and then use the output photocurrent to realize a real-time 
reduction of the mirror fluctuations.
The proposed scheme is a sort of continuous version of the 
stochastic cooling technique used in accelerators \cite{ACC}, 
because the feedback continuously
``kicks'' the mirror in order to put it in its
equilibrium position. This proposal has been experimentally realized in
Ref.~\cite{HEIPRL}, using the ``cold damping'' technique \cite{COLDD}, 
which amounts to applying a viscous feedback force to the oscillating mirror.
In the experimental studies of optomechanical systems performed
up to now, the effects of quantum noise
are blurred by thermal noise and the experimental results can be well
explained in classical terms (see for example \cite{PINARD}). 
However, developing 
a fully quantum description of the system in the presence of feedback is
of fundamental importance, for two main reasons.
First of all it allows to establish the conditions under which 
the effects of quantum noise in optomechanical systems become visible 
and experimentally detectable. 
We have recently shown in Ref.~\cite{LETTER}
that there is an appreciable difference
between the classical and quantum description of feedback already 
at liquid He temperatures.
Moreover, a completely quantum treatment allows to establish the ultimate
limits of the proposed feedback schemes, as for example, the 
possibility
to reach ground state cooling of a mechanical, macroscopic degree of 
freedom. In Ref.~\cite{MVTPRL}, a quantum treatment
of stochastic cooling feedback has been already presented, based however
on a master equation description which is not valid at very
low temperatures \cite{VICTOR}. A consistent quantum description of
both stochastic cooling and cold damping feedback schemes, valid at
all temperatures, has been presented in \cite{LETTER},
and recently a discussion of the quantum limits of cold damping has been
presented in \cite{COURTY}. The present paper will extend
and generalize the results of \cite{LETTER,COURTY},  
allowing us to make a detailed comparison of the two feedback schemes, 
and to establish all their potential applications.
In particular we shall see that both schemes can
achieve ground state cooling of an oscillating mode of the mirror, 
and that, in an appropriate limit, 
the ``stochastic cooling'' feedback of Ref.~\cite{MVTPRL}, can 
even break the standard quantum limit, achieving steady state 
position squeezing. The experimental realization of these quantum 
limits in optomechanical systems is extremely difficult, but the 
feedback methods described in this paper may be useful also
for microelectromechanical systems, where the search for quantum 
effects in mechanical systems is also very active \cite{BLENC,ROUK}.

Thermal noise reduction is important, but is not the only relevant 
aspect. What is more important, expecially for 
gravitational wave detection \cite{GRAV}, or for
metrology applications \cite{ROUK}, is to improve the sensitivity, 
i.e., the signal to noise ratio (SNR) of position measurements 
\cite{onofrio}.
Both the stochastic cooling scheme of Ref.~\cite{MVTPRL} and the cold 
damping scheme of Ref.~\cite{HEIPRL} cool the mirror by overdamping it, 
thereby strongly decreasing its mechanical susceptibility at resonance.
Cooling is therefore achieved through the 
suppression of the resonance peak in the noise power spectrum.
This suggests that both feedback schemes cannot be directly applied 
to improve the sensitivity for the detection of weak forces, because
the strong reduction of the mechanical susceptibility at resonance means that
the mirror does not respond both to the noise and to the signal.  
We shall see that this is true only in stationary conditions, i.e.,
we shall prove that the {\em stationary} spectral SNR
is never improved by feedback. However, as we have recently shown in
\cite{LETTER}, it is possible to use feedback
with an appropriate {\em nonstationary} strategy,
able to increase significantly the SNR for the detection
of {\em impulsive} classical forces acting on the oscillator.
Here we shall extend the results of \cite{LETTER}, by adopting
a general description of nonstationary spectral measurements.

The outline of the paper is as follows. In Sec. II 
we describe the model and derive the 
appropriate quantum Langevin equations. 
In Sec. III we describe the stochastic cooling feedback scheme of 
Ref.~\cite{MVTPRL} and the cold damping feedback using the 
quantum Langevin theory developed in \cite{HOW,GTV},
and we make a detailed comparison of the two schemes.
In Section IV we analyze the stationary state of the oscillating mirror, 
and we determine the conditions under which feedback can be used to
achieve ground state cooling or position squeezing.
In Section V we present a general description of nonstationary 
spectral measurement and we discuss the stationary limit in particular.
Section VI describes how the sensitivity
of position measurements can be improved by using feedback in a 
nonstationary way, and Section VII is for 
concluding remarks.

\section{The model}

\begin{figure}[t]
\centerline{\epsfig{figure=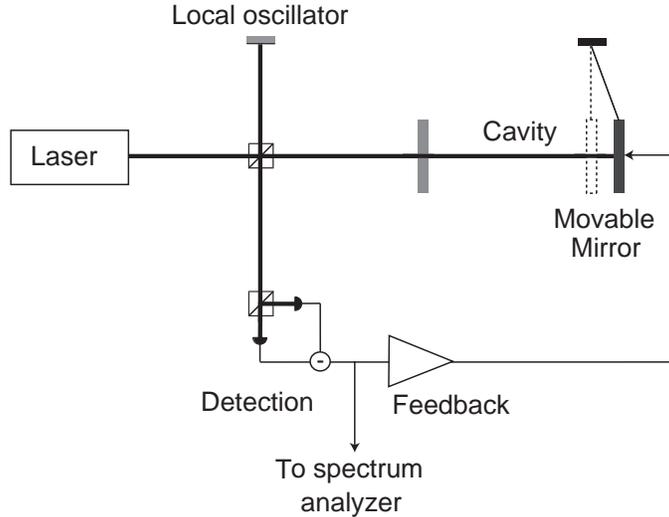,width=3.5in}}
\caption{\widetext 
Schematic description of the system.
The cavity mode is driven by the laser which,
thanks to the beam splitter, provides also the 
local oscillator for the homodyne measurement.
The signal is then fed back to the mirror motion.}
\label{fig1}
\end{figure}

The system studied in the present paper consists of a coherently 
driven optical cavity with a moving mirror (Fig.~\ref{fig1}). 
This opto-mechanical 
system can represent one arm of an interferometer able to detect weak 
forces as those associated with gravitational waves \cite{GRAV}, or 
an atomic force microscope \cite{AFM}. 
The detection 
of very weak forces requires having quantum limited devices, whose 
sensitivity is ultimately determined by the quantum fluctuations.
For this reason we shall describe the mirror as a single {\em quantum} 
mechanical harmonic oscillator with mass $m$ and frequency 
$\omega_{m}$. Experimentally, the mirror motion is the result of the 
excitation of many vibrational modes, including internal acoustic modes.
The description of the mirror as a single oscillator is however a good
approximation when frequencies are limited to a bandwidth including 
a single mechanical resonance, by using for example a bandpass filter
in the detection loop \cite{hadjar2}.

The optomechanical coupling between the mirror and
the cavity field is realized by the radiation pressure. The 
electromagnetic field exerts a force on the movable mirror which is 
proportional to the intensity of the field, which, at the same time, 
is phase-shifted by $2kq$, where $k$ is the wave vector and $q$ is
the mirror displacement from the equilibrium position.
In the adiabatic limit in which the mirror frequency is much smaller 
than the cavity free spectral range $c/2L$ ($L$ is the cavity length)
\cite{LAW}, one can focus on one cavity mode only because
photon scattering into other modes can be neglected, and one has the 
following Hamiltonian \cite{HAMI}
\begin{equation}
	H=\hbar \omega_{c}b^{\dagger}b + 
    \hbar\omega_m\left(P^{2}+
	Q^{2}\right) 
	-\hbar G
	b^{\dagger}b Q + i\hbar E\left(b^{\dagger}e^{-i\omega_{0}t}-b
	e^{i\omega_{0}t}\right) \,, 
	\label{HINI}
\end{equation}
where $b$ is the cavity mode annihilation operator with optical
frequency $\omega_{c}$, and $E$ describes the coherent input field
with frequency $\omega_{0}\sim \omega_{c}$
driving the cavity. 
Moreover, $Q$ and $P$ are the dimensionless position and momentum
operator of the movable mirror, with $\left[Q,P\right]=i/2$, 
and $G=(\omega_c/L)\sqrt{\hbar/2m\omega_m}$ is the coupling constant.
Since we shall focus on the quantum and thermal noise of the system,
we shall neglect all the technical sources of noise, i.e., 
we shall assume that the driving laser is stabilized
in intensity and frequency. This means neglecting all
the fluctuations of the complex parameter $E$. Including these 
supplementary noise sources is however quite straightforward and a 
detailed calculation of their effect is shown in Ref. \cite{KURT}. 
Moreover recent experiments have shown
that classical laser noise can be made negligible in the relevant 
frequency range \cite{HADJAR,TITTO}.
The adiabatic regime $\omega_{m} \ll c/2L$ we have assumed 
in Eq.~(\ref{HINI}) implies $\omega_{m} \ll \omega_{c}$,
and therefore the generation of photons due to the Casimir effect,
and also retardation and Doppler effects are completely negligible.

The dynamics of the system is not only determined by the Hamiltonian
interaction (\ref{HINI}), but also by the dissipative interaction
with external degrees of freedom. The cavity mode is damped 
due to the
photon leakage through the mirrors which couple the cavity 
mode with the continuum of the outside electromagnetic modes. 
For simplicity 
we assume that the movable mirror has perfect reflectivity and that
transmission takes place through the other, ``fixed'', mirror only.
We indicate the photon decay rate at the fixed mirror by $\gamma_c$.
Then, the quantity $E$ is related to the input laser 
power $\wp$ by $E=\sqrt{\wp\gamma_{c}/\hbar \omega_{0}}$.
The mechanical oscillator, which may represent not only the
center-of-mass degree of freedom of the mirror, but also a torsional 
degree of freedom as in \cite{TITTO}, or an internal acoustic mode
as in \cite{HADJAR}, undergoes Brownian motion caused by the 
uncontrolled coupling with other internal and external modes at 
thermal equilibrium.

The dynamics of the system
can be described by the following set of coupled quantum 
Langevin equations (QLE)
(in the interaction picture with respect to 
$\hbar \omega_{0}b^{\dagger} b$) 
\begin{mathletters}\label{QLENL}
\begin{eqnarray}
\dot{Q}(t) &=& \omega_m P(t) \,, 
\label{QLENL1}\\
\dot{P}(t) &=& -\omega_{m} Q(t) + {\cal W}(t) -  
 {\gamma_m} P(t)   
 +G b^{\dagger}(t) b(t) \,,
\label{QLENL2}\\
\dot{b}(t) &=& - \left(i \omega_{c} - i \omega_{0} +
\frac{\gamma_{c}}{2}\right) b(t) + 2 i 
G Q(t) b(t) + E +
\sqrt{\gamma_{c}}b_{in}(t)\,, 
\label{QLENL3}
\end{eqnarray}
\end{mathletters}
where $b_{in}(t)$ is the input noise operator \cite{GAR} 
associated with the vacuum 
fluctuations of the continuum of modes 
outside the cavity, having the 
following correlation functions 
\begin{mathletters}\label{INCOR}
\begin{eqnarray}
&& \langle b_{in}(t)b_{in}(t') \rangle = \langle 
b_{in}^{\dagger}(t)b_{in}(t') \rangle
= 0 \,,
\label{INCOR1}\\
&& \langle b_{in}(t)b_{in}^{\dagger}(t') \rangle = \delta(t-t') \,.
\label{INCOR2} 
\end{eqnarray}
\end{mathletters}
Furthermore, ${\cal W}(t)$ is the Brownian noise operator
defined consistently with quantum mechanics \cite{VICTOR}.
It has the following correlation functions
\begin{equation}
\langle {\cal W}(t) {\cal W}(t^\prime) \rangle=
\frac{1}{2\pi}\frac{\gamma_m}{\omega_m} 
\Big\{ {\cal F}_{r}(t-t^\prime) + i 
{\cal F}_{i}(t-t^\prime) \Big\}\,,
\label{BROWCOR}
\end{equation}
where
\begin{mathletters}\label{F}
\begin{eqnarray}
{\cal F}_{r}(t)&=&\int_{0}^{\varpi} d\omega \;
  \omega \cos(\omega t) \coth\left(\frac{\hbar\omega}{2 k_B T} 
  \right) \,,
  \label{FR} \\
{\cal F}_{i}(t)&=& - \int_{0}^{\varpi} d\omega \;
  \omega \sin(\omega t) \,,
  \label{FI}  
\end{eqnarray}
\end{mathletters}
with $T$ the bath temperature, $\gamma_m$ the mechanical 
decay rate, $k_B$ the 
Boltzmann constant, and $\varpi$ the frequency cutoff of the reservoir
spectrum.
The antisymmetric part, ${\cal F}_{i}$, of Eq.~(\ref{BROWCOR}), 
is a direct
consequence of the commutation relations for the Brownian noise 
operator, and the symmetric part, ${\cal F}_{r}$
explicitely depends on temperature 
and becomes proportional to a Dirac delta function when the 
high temperature limit $ k_BT \gg \hbar \varpi $ first, 
and the infinite frequency 
cutoff limit $\varpi \rightarrow \infty$ later, are taken. 
Eqs.~(\ref{BROWCOR}) and (\ref{F})
show the non-Markovian nature of quantum Brownian motion, which 
becomes particularly evident in the low temperature limit 
\cite{GRAB,HAAKE}. Therefore, the {\em exact} QLE
(\ref{QLENL}) reduce to the standard ones
\cite{GAR} in the limit $\varpi \rightarrow \infty$.
It is also important to stress that 
the quantum Langevin description of quantum Brownian 
motion given by Eq.~(\ref{QLENL}) is more general than 
that associated with a master equation 
approach, because it is valid {\em at all temperatures} 
and it does not need any high temperature limit \cite{VICTOR}. 

In standard interferometric applications, the driving field is
very intense. Under this condition, the system is characterized by a
semiclassical steady state with 
the internal cavity mode in a coherent
state $|\beta\rangle $, and a new equilibrium position for the 
mirror, displaced by $G| \beta |^{2}/\omega_m$ 
with respect to that with no driving field.
The steady state amplitude is given by the solution
of the nonlinear equation
\begin{equation}
	\beta=\frac{E}{\frac{\gamma_{c}}{2}+i \omega_{c} - i \omega_{0}
	-2 i \frac{G^2}{\omega_{m}}
        |\beta|^{2}} ,
	\label{BETA}
\end{equation}
which is obtained by taking the expectation values of 
Eqs.~(\ref{QLENL}), factorizing them and setting 
all the time derivatives to zero. Eq.~(\ref{BETA}) shows a bistable 
behaviour which has been experimentally observed in \cite{DORSEL}.
Under these semiclassical conditions, the dynamics is well described 
by linearizing the QLE 
(\ref{QLENL}) around the steady state. If we now
rename with $Q(t)$ and $b(t)$ the operators describing the quantum 
fluctuations around the classical steady state, we get
\begin{mathletters}\label{QLE}
\begin{eqnarray}
\dot{Q}(t) &=& \omega_m P(t) \,, 
\label{QLEL1}\\
\dot{P}(t) &=& -\omega_{m} Q(t) -  
  {\gamma_m} P(t)   +G\beta
  \left[ b(t) + 
  b^{\dagger}(t) \right] + {\cal W}(t) \,,
\label{QLEL2}\\
\dot{b}(t) &=&  -\left(\frac{\gamma_{c}}{2} + i \Delta\right)b(t)
+2 i G \beta Q(t)+ \sqrt{\gamma_{c}} b_{in}(t) \,,
\label{QLEL3}
\end{eqnarray} 
\end{mathletters}
where we have chosen the phase of the cavity mode field so that
$\beta$ is real and 
\begin{equation}
	\Delta = \omega_{c} - \omega_{0} - 
	\frac{2G^2}{\omega_m} \beta^{2},
	\label{DETU}
\end{equation}
is the cavity mode detuning. 
We shall consider from now on $\Delta=0$, 
which corresponds to the most common experimental situation, 
and which can
always be achieved by appropriately adjusting the driving
field frequency $\omega_{0}$. In this case the dynamics becomes 
simpler, and, introducing the field phase quadrature 
$Y(t)=i\left(b^{\dagger}(t)-b(t)\right)/2$ and field amplitude 
quadrature $X(t)=\left(b(t)+b^{\dagger}(t)\right)/2$, 
one has that only the phase quadrature $Y(t)$ is 
affected by the mirror position fluctuations $Q(t)$,
while the amplitude field quadrature $X(t)$ is not. In fact, the 
linearized QLE (\ref{QLE}) can be rewritten as
\begin{mathletters}\label{QLE2}
\begin{eqnarray}
\dot{Q}(t) &=& \omega_m P(t) \,, 
\label{QLE2L1}\\
\dot{P}(t) &=& -\omega_{m} Q(t) -  
  {\gamma_m} P(t)   +2 G\beta X(t) + {\cal W}(t) \,,
\label{QLE2L2}\\
\dot{Y}(t) &=&  -\frac{\gamma_{c}}{2} Y(t)
+2 G \beta Q(t)+ \frac{\sqrt{\gamma_{c}}}{2} Y_{in}(t) \,,
\label{QLE2L3} \\
\dot{X}(t) &=&  -\frac{\gamma_{c}}{2} X(t)
+ \frac{\sqrt{\gamma_{c}}}{2} X_{in}(t) \,,
\label{QLE2L4}
\end{eqnarray} 
\end{mathletters}
where we have introduced the phase input noise 
$Y_{in}(t)=i\left(b_{in}^{\dagger}(t)-b_{in}(t)\right)$ and the 
amplitude input noise $X_{in}(t)=b_{in}^{\dagger}(t)+b_{in}(t)$.

\section{Position measurement and feedback}

Usually the movable mirror is used as a ponderomotive meter to detect 
small forces acting on it \cite{BK}. Thus, we introduce an 
additional Hamiltonian term describing the action of a classical
external force $f(t)$, that is
\begin{equation}\label{HEXT}
H_{ext}=-Q f(t)\,.
\end{equation}
Information about such a force can be obtained 
by looking at the mechanical oscillator 
position $Q(t)$. The position measurement is commonly performed in the 
large cavity bandwidth limit $\gamma_{c} \gg G \beta $, $\omega_m$,
when the 
cavity mode dynamics adiabatically follows that of the movable mirror
and it can be eliminated, that is, from Eq.~(\ref{QLE2L3}),
\begin{equation}
Y(t) \simeq \frac{4G \beta}{\gamma_{c}}Q(t) 
+\frac{Y_{in}(t)}{\sqrt{\gamma_{c}}},
\label{adiab}
\end{equation}
and $X(t) \simeq X_{in}(t)/\sqrt{\gamma_{c}}$ from Eq.~(\ref{QLE2L4}).
Performing a continuous 
homodyne measurement of the phase quadrature $Y(t)$ means 
therefore continuously monitoring the real time dynamics of the 
oscillator position $Q(t)$, which, in turn, implies detecting the 
effects of classical force $f(t)$.
The experimentally detected quantity is the output homodyne photocurrent
\cite{HOW,GTV,homo}
\begin{equation}\label{BOUNDARY}
Y_{out}(t)=2\eta \sqrt{\gamma_{c}}Y(t)-\sqrt{\eta}Y_{in}^{\eta}(t)\,,
\end{equation}
where $\eta$ is the detection efficiency and $Y_{in}^{\eta}(t)$
is a {\em generalized phase input noise}, coinciding with the
input noise $Y_{in}(t)$ in the case of perfect detection $\eta 
=1$, and taking into account the additional noise due to the 
inefficient detection in the general case $\eta < 1$ \cite{GTV}.
This generalized phase input noise can be written in terms 
of a generalized input noise $b_{\eta}(t)$ as
$Y_{in}^{\eta}(t)=i\left[b_{\eta}^{\dagger}(t)-b_{\eta}(t)\right]$. 
The quantum noise $b_{\eta}(t)$ is correlated with the 
input noise $b_{in}(t)$ and it is characterized by the following 
correlation functions \cite{GTV}
\begin{mathletters}\label{INCORETA}
\begin{eqnarray}
&& \langle b_{\eta}(t)b_{\eta}(t') \rangle = \langle 
b_{\eta}^{\dagger}(t)b_{\eta}(t') \rangle
= 0 \,,
\label{INCORETA1}\\
&& \langle b_{\eta}(t)b_{\eta}^{\dagger}(t') \rangle = \delta(t-t') 
\,,
\label{INCORETA2} \\
&& \langle b_{in}(t)b_{\eta}^{\dagger}(t')
\rangle = \langle b_{\eta}(t)b_{in}^{\dagger}(t')
\rangle = \sqrt{\eta}\delta(t-t').
\label{INCORETA3}  
\end{eqnarray}
\end{mathletters}
The output of the homodyne measurement may be used to devise a
phase-sensitive feedback loop to control the dynamics of the mirror.
For example, we have proposed in Ref.~\cite{MVTPRL} to reduce 
the effects of thermal noise on the mirror
by feeding back the output homodyne photocurrent in an appropriate way.
The proposed scheme is a sort of continuous version of the 
stochastic cooling technique used in accelerators \cite{ACC}, 
because the homodyne 
measurement provides a continuous monitoring of the oscillator's
position, and the feedback continuously
``kicks'' the mirror in order to put it in its
equilibrium position. Our proposal of cooling the mirror using a 
feedback loop has been experimentally realized in
Ref.~\cite{HEIPRL} (see also \cite{PINARD}), using a different method,
the so-called ``cold damping'' technique \cite{COLDD}. 
This latter feedback scheme shares some analogies with that proposed 
in Ref.~\cite{MVTPRL} and 
amounts to applying a viscous feedback force to the oscillating mirror.
In the experiment of Refs.~\cite{HEIPRL,PINARD}, the viscous force is provided
by the radiation pressure of another laser beam, intensity-modulated by the
time derivative of the homodyne signal.

The effect of the feedback loop has been 
described using quantum trajectory theory \cite{howmil} and the master 
equation formalism in Ref.~\cite{MVTPRL}, and a classical 
description neglecting all quantum fluctuations
in Ref.~\cite{HEIPRL,PINARD}. Here we shall use a more general 
description of feedback based on QLEs for
Heisenberg operators, first developed in Ref.~\cite{HOW} and 
generalized to the non-ideal detection case in Ref.~\cite{GTV}
(see also \cite{DOHERTY} for a comparison between these
quantum feedback approaches and general quantum control theories).
This general quantum description of feedback will allow us to compare 
the two different feedback 
schemes, the stochastic cooling scheme of Ref.~\cite{MVTPRL}, and the 
cold damping scheme of Ref.~\cite{HEIPRL,PINARD}. Moreover,
the present quantum treatment will allow us to show
that in the presence of feedback
the radiation quantum noise has important effects, and 
that a classical stochastic treatment of the dynamics
of the system is generally inadequate.
Our treatment explicitely includes the limitations due to the 
quantum efficiency of the detection, but neglects other
possible technical imperfections of the feedback loop, as for
example the electronic noise of the feedback loop (discussed
in \cite{PINARD}), or the fluctuations of the laser beam used for
the feedback in the cold damping scheme.

\subsection{Stochastic cooling}

Let us first consider the stochastic cooling scheme of
Ref.~\cite{MVTPRL}. In this scheme, the feedback loop induces a 
continuous position shift controlled by the output homodyne 
photocurrent $Y_{out}(t)$.
This effect of feedback manifests itself in an additional term
in the QLE for a generic operator ${\cal O}(t)$ given by \cite{GTV}
\begin{equation}\label{DOSTRA}
\dot{{\cal O}}_{fb}(t)=
i\frac{\sqrt{\gamma_{c}}}{\eta}
Y_{out}(t-\tau)
\left[g_{sc}P(t),{\cal O}(t)\right]\,,
\end{equation}
where $\tau$ is the feedback loop delay time, and $g_{sc}$ is a 
dimensionless feedback 
gain factor. 
The feedback delay-time is essentially determined by the electronics
involved in the feedback loop and is always much smaller than the typical 
timescale of the mirror dynamics. It is therefore common to consider
the zero delay-time limit, $\tau \to 0$. This limit is however quite delicate
in general \cite{HOW,GTV}. In fact, $Y_{out}(t-\tau)$, being an output operator, 
commutes with $\left[g_{sc}P(t),{\cal O}(t)\right]$ for any nonzero 
$\tau$, but this is no 
more true when $\tau = 0$. Therefore, one has to be 
careful with ordering in the zero delay-time limit. However, with the choice
of Eq.~(\ref{DOSTRA}) for the feedback term, the only nonzero commutator
in the QLE of Eq.~(\ref{QLE2}) is $\left[g_{sc}P(t),Q(t)\right]$, 
which, being a c-number, does not
create any ordering ambiguity. Therefore one has exactly the same equations
one would have by putting
directly $\tau=0$ in Eq.~(\ref{DOSTRA}), that is,
\begin{mathletters}\label{QFBEQ}
\begin{eqnarray}
\dot{Q}(t)&=&\omega_{m}P(t)
+g_{sc}\gamma_{c}Y(t)-\frac{g_{sc}}{2}\sqrt{
\frac{\gamma_c}{\eta}}Y_{in}^{\eta}(t)\,,
\label{QFBEQ1}\\
\dot{P}(t) &=& -\omega_{m} Q(t) -  
  {\gamma_m} P(t)   +2 G\beta X(t) + {\cal W}(t) +f(t)\,,
\label{QFBEQ2}\\
\dot{Y}(t) &=&  -\frac{\gamma_{c}}{2} Y(t)
+2 G \beta Q(t)+ \frac{\sqrt{\gamma_{c}}}{2} Y_{in}(t) \,,
\label{QFBEQ3} \\
\dot{X}(t) &=&  -\frac{\gamma_{c}}{2} X(t)
+ \frac{\sqrt{\gamma_{c}}}{2} X_{in}(t) \,,
\label{QFBEQ4}
\end{eqnarray} 
\end{mathletters}
where we have used Eq.~(\ref{BOUNDARY}).
After the adiabatic elimination of the radiation mode (see 
Eq.~(\ref{adiab})), the above
equations reduce to
\begin{mathletters}\label{QEQ}
\begin{eqnarray}
\dot{Q}(t)&=&\omega_{m}P(t)
+4 G \beta g_{sc} Q(t)+\sqrt{\gamma_{c}}g_{sc}
Y_{in}(t)-\frac{g_{sc}}{2}
\sqrt{\frac{\gamma_{c}}{\eta}}Y_{in}^{\eta}(t)\,,
\label{QEQ1}\\
\dot{P}(t)&=&-\omega_{m}Q(t)
-\gamma_{m}P(t)+\frac{2 G \beta}{\sqrt{\gamma_{c}}}
X_{in}(t)+{\cal W}(t)+f(t)\,.
\label{QEQ2}
\end{eqnarray}
\end{mathletters}
The solution of these QLE
for the conjugate operators $Q(t)$ and $P(t)$ 
can be easily obtained by performing the Laplace transform,
and they will be useful in the following.
Their expression is
\begin{mathletters}\label{TIME}
\begin{eqnarray}
Q(t)&=&K_Q(t) Q(0)+\chi_{sc}(t) P(0)
+\int_{0}^{t}\,dt'\,\chi_{sc}(t')f(t-t')
\nonumber\\
&&+\int_{0}^{t}\,dt'\,K_Q(t')\left[
\sqrt{\gamma_{c}}g_{sc}
Y_{in}(t-t')-\frac{g_{sc}}{2}
\sqrt{\frac{\gamma_{c}}{\eta}}Y_{in}^{\eta}(t-t')\right]
\nonumber\\
&&+\int_{0}^{t}\,dt'\,\chi_{sc}(t')\left[
{\cal W}(t-t')+\frac{2 G \beta}{\sqrt{\gamma_{c}}}
X_{in}(t-t')\right]\,,\label{QTIME} \\
P(t)&=&K_P(t) P(0)-\chi_{sc}(t) Q(0)
+\int_{0}^{t}\,dt'\,K_P(t')f(t-t')
\nonumber\\
&&-\int_{0}^{t}\,dt'\,\chi_{sc}(t') \left[
\sqrt{\gamma_{c}}g_{sc}
Y_{in}(t-t')-\frac{g_{sc}}{2}
\sqrt{\frac{\gamma_{c}}{\eta}}Y_{in}^{\eta}(t-t')\right]
\nonumber\\
&&+\int_{0}^{t}\,dt'\,K_P(t')\left[
{\cal W}(t-t')+\frac{2 G \beta}{\sqrt{\gamma_{c}}}
X_{in}(t-t')\right] \label{PTIME}\,.
\end{eqnarray}
\end{mathletters}
We have introduced the time-dependent susceptibility 
$\chi_{sc}(t)$ describing the response of the movable mirror
in the presence of the stochastic cooling feedback
\begin{equation}\label{CHITILSC}
\chi_{sc}(t)=\frac{\omega_{m}}{\sqrt{\omega_m^{2}-\gamma_{m}^{2}\left(
\frac{1-g_{1}}{2}\right)^{2}}}
e^{-(1+g_{1})\gamma_{m}t/2}
\sin\left[t\sqrt{\omega_m^{2}-\gamma_{m}^{2}\left(
\frac{1-g_{1}}{2}\right)^{2}}\right],
\end{equation}
and the two related response functions
\begin{eqnarray}\label{KQ}
&&K_Q(t)=\frac{\dot{\chi}_{sc}(t)+\gamma_m
\chi_{sc}(t)}{\omega_m}, \\
&&K_P(t)=\frac{\dot{\chi}_{sc}(t)+g_1
 \chi_{sc}(t)}{\omega_m} . \label{KP}
\end{eqnarray}
We have also rescaled the feedback gain and defined
$g_1 = -4G\beta g_{sc}/\gamma_{m}$.

\subsection{Cold damping}

Cold damping techniques, that is, the possibility to use a feedback loop
to reduce the effective 
temperature of a system well below the operating temperature,
have been applied in classical electromechanical systems
for many years \cite{COLDD}, 
and only recently they have been proposed to improve
cooling and sensitivity at the quantum level \cite{GRAS}.
This technique is based on the application of a negative derivative
feedback, which increases the damping of the system without 
correspondingly increasing the thermal noise \cite{COLDD,GRAS}.
This technique has been succesfully applied for the first
time to an optomechanical 
system composed of a high-finesse cavity with a movable mirror in 
the experiments of Refs.~\cite{HEIPRL,PINARD}. 
In these experiments, the displacement of the mirror is measured with 
very high sensitivity \cite{HADJAR}, and the obtained information
is fed back to the 
mirror via the radiation pressure of another, intensity-modulated, laser
beam, incident on the back of the mirror.
Cold damping is obtained by modulating with the {\em time derivative}
of the homodyne signal, in such a way that
the radiation 
pressure force is proportional to the mirror velocity.
The servo-control force then corresponds to a viscous force.
The results of Refs.~\cite{HEIPRL,PINARD} referred to a room temperature
experiment, and have been explained using a classical description.
The quantum description of cold damping in this optomechanical system
has been presented in \cite{LETTER} (see also Ref.~\cite{GRAS})),
and we shall follow this treatment.

In the quantum Langevin description, cold damping feedback scheme
implies the following additional term
in the QLE for a generic operator ${\cal O}(t)$ \cite{LETTER},
\begin{equation}\label{DLORO}
\dot{{\cal O}}_{fb}(t)=
\frac{i}{\eta \sqrt{\gamma_{c}}}
\dot{Y}_{out}(t-\tau)
\left[g_{cd}Q(t),{\cal O}(t)\right]\,.
\end{equation}
As for the stochastic cooling feedback case, one has only one
nonzero feedback term in the QLE of the system (\ref{QLE2}), which in this
case is $\left[g_{cd}Q(t),P(t)\right]$. Since this commutator is a c-number 
also in this case, we do not have any ordering problem 
in the zero delay-time limit,
and the QLE for the cold damping feedback scheme become
\begin{mathletters}\label{CFBEQ}
\begin{eqnarray}
\dot{Q}(t)&=&\omega_{m}P(t) \,,
\label{CFBEQ1}\\
\dot{P}(t) &=& -\omega_{m} Q(t) -  
  {\gamma_m} P(t)   +2 G\beta X(t)  
  -g_{cd}\dot{Y}(t)+\frac{g_{cd}}{2\sqrt{\gamma_c \eta}}
  \dot{Y}_{in}^{\eta}(t)
+{\cal W}(t) +f(t) \,,
\label{CFBEQ2}\\
\dot{Y}(t) &=&  -\frac{\gamma_{c}}{2} Y(t)
+2 G \beta Q(t)+ \frac{\sqrt{\gamma_{c}}}{2} Y_{in}(t) \,,
\label{CFBEQ3} \\
\dot{X}(t) &=&  -\frac{\gamma_{c}}{2} X(t)
+ \frac{\sqrt{\gamma_{c}}}{2} X_{in}(t) \,.
\label{CFBEQ4}
\end{eqnarray} 
\end{mathletters}
Adiabatically 
eliminating the cavity mode, one has
\begin{mathletters}\label{QLEFCDAD}
\begin{eqnarray}
\dot{Q}(t) &=& \omega_m P(t), 
 \label{QLEFCDAD1}\\
\dot{P}(t) &=& -\omega_{m} Q(t) -  
  {\gamma_m} P(t)   +\frac{2G\beta }{\sqrt{\gamma_{c}}}
  X_{in}(t) + {\cal W}(t) +f(t) - \frac{4G \beta g_{cd}}
 {\gamma_{c}}\dot{Q}(t)-\frac{g_{cd}}{\sqrt{\gamma_{c}}}\dot{Y}_{in}(t)
 + \frac{g_{cd}}{2\sqrt{\gamma_{c}\eta}}\dot{Y}_{in}^{\eta}(t).
\label{QLEFCDAD2}
\end{eqnarray}
\end{mathletters}
Notice that the modulation with the derivative of the homodyne photocurrent
implies the introduction of two new quantum input noises,
$\dot{Y}_{in}(t)$ and $\dot{Y}_{in}^{\eta}(t)$, whose correlation functions
can be simply obtained by differentiating the corresponding correlation
functions of $Y_{in}(t)$ and $Y_{in}^{\eta}(t)$. We have therefore
\begin{mathletters} \label{dotcorre}
\begin{eqnarray}
\langle \dot{Y}_{in}(t) \dot{Y}_{in}(t')\rangle &=&
\langle \dot{Y}_{in}(t') \dot{Y}_{in}(t)\rangle = 
\langle \dot{Y}_{in}^{\eta}(t) \dot{Y}_{in}^{\eta}(t')\rangle =
\langle \dot{Y}_{in}^{\eta}(t') \dot{Y}_{in}^{\eta}(t)\rangle
=-\ddot{\delta}(t-t'), \\
 \langle \dot{Y}_{in}^{\eta}(t) \dot{Y}_{in}(t')\rangle &=&
\langle \dot{Y}_{in}(t') \dot{Y}_{in}^{\eta}(t)\rangle
=-\sqrt{\eta}\ddot{\delta}(t-t'), \\
\langle X_{in}(t) \dot{Y}_{in}^{\eta}(t')\rangle &=&
-\langle \dot{Y}_{in}^{\eta}(t') X_{in}(t)\rangle
=-i\sqrt{\eta}\dot{\delta}(t-t').
\end{eqnarray}
\end{mathletters}
In this case the solution of the
adiabatic QLE reads
\begin{eqnarray}\label{QT}
Q(t)&=&K(t) Q(0)+\chi_{cd}(t) P(0)
+\int_{0}^{t}\,dt'\,\chi_{cd}(t')f(t-t')
\nonumber\\
&&+\int_{0}^{t}\,dt'\,\chi_{cd}(t-t')\left[
\frac{2G\beta }{\sqrt{\gamma_{c}}}
  X_{in}(t') + {\cal W}(t')
  -\frac{g_{cd}}{\sqrt{\gamma_{c}}}\dot{Y}_{in}(t')
 + \frac{g_{cd}}{2\sqrt{\gamma_{c}\eta}}\dot{Y}_{in}^{\eta}(t')
\right]\,,
\end{eqnarray} 
and
$P(t) = \dot{Q}(t)/\omega_{m}$,
where we have introduced the time-dependent susceptibility
in the case of the cold damping feedback scheme
\begin{equation}\label{CHITILCD}
\chi_{cd}(t)=\frac{\omega_{m}}{\sqrt{\omega_m^{2}-\gamma_{m}^{2}
\left(\frac{1+g_{2}}{2}\right)^{2}}}
e^{-(1+g_{2})\gamma_{m}t/2}
\sin\left[t\sqrt{\omega_m^{2}-\gamma_{m}^{2}\left(
\frac{1+g_{2}}{2}\right)^{2}}\right]
\end{equation}
and the related response function 
\begin{equation} \label{K}
K(t) = 1- \omega_{m}\int_{0}^{t}dt'\,\chi_{cd}(t').
\end{equation}
We have again rescaled the feedback gain and defined
$g_{2}=4 G \beta \omega_{m}g_{cd}/\gamma_{m}\gamma_{c}$. 

\subsection{Comparison between the two feedback schemes}

The two sets of QLE for the mirror Heisenberg operators,
Eqs.~(\ref{QEQ}) and (\ref{QLEFCDAD}), show that the two feedback 
schemes are not exactly equivalent. They are however physically 
analogous, as it can be seen, for example, by looking at the 
differential equation for the displacement operator $Q(t)$. 
In fact, from Eqs.~(\ref{QEQ}) one gets 
\begin{eqnarray}
\ddot{Q}(t)+\left(1+g_{1}\right)\gamma_{m}\dot{Q}(t) 
+\left(\omega_{m}^{2}+\gamma_{m}^{2}g_{1}\right)Q(t)&=&\omega_{m}\left[
\frac{2 G \beta}{\sqrt{\gamma_{c}}}X_{in}(t)+{\cal W}(t)+f(t)\right]
+\sqrt{\gamma_{c}}g_{sc}
\dot{Y}_{in}(t)-\frac{g_{sc}}{2}
\sqrt{\frac{\gamma_{c}}{\eta}}\dot{Y}_{in}^{\eta}(t) \nonumber \\
&+&
\frac{\gamma_{m}}{\omega_{m}}\left[\sqrt{\gamma_{c}}g_{sc}
Y_{in}(t)-\frac{g_{sc}}{2}
\sqrt{\frac{\gamma_{c}}{\eta}}Y_{in}^{\eta}(t)\right],
\end{eqnarray}
for the stochastic cooling scheme,
while from Eqs.~(\ref{QLEFCDAD}) one gets
\begin{equation}
\ddot{Q}(t)+\left(1+g_{2}\right)\gamma_{m}\dot{Q}(t) +
\omega_{m}^{2}Q(t)=\omega_{m}\left[\frac{2G\beta }{\sqrt{\gamma_{c}}}
  X_{in}(t) + {\cal W}(t) +f(t) -\frac{g_{cd}}{\sqrt{\gamma_{c}}}\dot{Y}_{in}(t)
 + \frac{g_{cd}}{2\sqrt{\gamma_{c}\eta}}\dot{Y}_{in}^{\eta}(t)
\right],
\end{equation}
for the cold damping scheme. 
These equations shows that in both schemes the main effect of 
feedback is the modification of mechanical damping $\gamma_{m} 
\rightarrow \gamma_{m}(1+g_{i})$ ($i=1,2$). In the stochastic cooling scheme one 
has also a frequency renormalization $\omega_{m}^{2} \rightarrow  
\omega_{m}^{2}+\gamma_{m}^{2}g_{1}$, which is however usually negligible 
since the mechanical quality factor ${\cal Q}=\omega_{m}/\gamma_{m}$
is always large. Moreover, in the two cases the 
position dynamics is affected by similar,
even though not identical, noise terms. This comparison shows that the 
stochastic cooling scheme of Ref.~\cite{MVTPRL} is also able to provide a cold 
damping effect of increased damping without an increased temperature
\cite{LETTER}.

\section{Stationary state and cooling}

We now study 
the stationary state of the movable mirror in the presence 
of the two feedback schemes, which is obtained by considering the 
dynamics in the asymptotic limit $t \to \infty$. We shall see 
that both feedback schemes are able
to lower the effective temperature of the system, and that, in 
particular limits, the steady state can have interesting quantum 
features. In fact, both schemes are able to achieve 
ground state cooling, and the stochastic cooling feedback is even able 
to achieve steady state position squeezing.

\subsection{Stochastic cooling feedback}

Using the solution (\ref{QTIME}), one has
\begin{equation} 
\langle Q^{2}\rangle _{st} = \lim_{t \to \infty}
\langle Q(t)^{2}\rangle = \int_{0}^{\infty}dt' \int_{0}^{\infty}dt''
K_{Q}(t')K_{Q}(t'') c_{1}(t'-t'') +
\int_{0}^{\infty}dt' \int_{0}^{\infty}dt''
\chi_{sc}(t')\chi_{sc}(t'') c_{2}(t'-t''), \label{q21}
\end{equation}
where $c_{1}(t)$ is the stationary symmetrized
correlation function of the noise 
term $n_{1}(t)=\sqrt{\gamma_{c}}g_{sc}
Y_{in}(t)-\frac{g_{sc}}{2}
\sqrt{\frac{\gamma_{c}}{\eta}}Y_{in}^{\eta}(t)$, 
$c_{2}(t)$ is the stationary symmetrized correlation function of the noise 
term $n_{2}(t)={\cal W}(t)+\frac{2 G \beta}{\sqrt{\gamma_{c}}}
X_{in}(t)$, and we have used the fact that $n_{1}(t)$ and 
$n_{2}(t)$ are uncorrelated. Using the correlation functions 
(\ref{INCOR}), (\ref{BROWCOR}), and (\ref{INCORETA}), one gets
\begin{mathletters}\label{corsc}
\begin{eqnarray}
c_{1}(t) &=& 
\frac{\gamma_{c}\gamma_{m}^{2}g_{1}^{2}}{64 \eta G^{2}\beta ^{2}} \delta (t)
 \label{corsc1}\\
c_{2}(t) &=& \frac{4 G^{2}\beta ^{2}}{\gamma_{c}} \delta (t) +
\frac{\gamma_{m}}{2 \pi \omega _{m}}{\cal F}_{r}(t).
 \label{corsc2} 
\end{eqnarray}
\end{mathletters}
The expression for $\langle Q^{2}\rangle _{st} $ is obtained 
using
Eqs.~(\ref{CHITILSC}), (\ref{KQ}), and (\ref{corsc})
in Eq.~(\ref{q21}),
\begin{equation}
\langle Q^{2}\rangle _{st} = 
\frac{\gamma_{c}\gamma_{m}g_{1}^{2}}{128 \eta G^{2}\beta ^{2}} 
\frac{1+{\cal Q}^{2}+g_{1}}{\left(1
+g_{1}\right)\left({\cal Q}^{2}+g_{1}\right)}+
\frac{2 G^{2}\beta ^{2}}{\gamma_{c}\gamma_{m}}
\frac{{\cal Q}^{2}}{\left(1+g_{1}\right)
\left({\cal Q}^{2}+g_{1}\right)} +
\langle Q^{2}\rangle _{BM}.
\label{q2sc}
\end{equation}
The term $\langle Q^{2}\rangle _{BM}$ is the 
contribution of the mirror quantum Brownian motion, whose general 
expression is obtained
by rewriting ${\cal F}_{r}(t'-t'')$ in Eq.~(\ref{q21}) 
in terms of its Fourier transform ${\cal F}_{r}(\omega )$
(see Eq.~(\ref{BROWCOR})), to get
\begin{equation}
\langle Q^{2}\rangle _{BM} = \int_{-\varpi}^{\varpi}\frac{d \omega}{2 
\pi} \frac{\gamma_{m}}{2 \omega_{m}}\omega \coth
\left(\frac{\hbar \omega}{2 k_{B} T}\right) \left 
|{\tilde \chi}_{sc}(\omega)\right|^{2},
\label{qbm}
\end{equation}
where 
\begin{equation}
{\tilde \chi}_{sc}(\omega)=\frac{\omega_{m}}{\omega_{m}^{2}+
	g_{1}\gamma_{m}^{2}-\omega^{2}+i\omega \gamma_{m}\left(1+g_{1}\right)}
	\label{suscsc}
\end{equation}
is the frequency-dependent susceptibility of the mirror in the 
stochastic cooling feedback scheme. The general analytical expression
of the quantum Brownian motion term $\langle Q^{2}\rangle _{BM}$, 
valid in any range of parameters, is cumbersome 
and has been obtained in \cite{GRAB,HAAKE}. However, in typical
optomechanical experiments \cite{HADJAR,TITTO,HEIPRL,PINARD} 
it is always $\hbar \gamma_{m} \ll \hbar \omega_{m} \ll k_B T$, 
and it is possible to see 
\cite{GRAB} that, in this limiting case, the classical approximation 
$\coth(\hbar \omega/2k_B T) \simeq 2k_B T/\hbar \omega $
(which is equivalent to approximate ${\cal F}_{r}(t) \simeq 
\left(\gamma_{m}k_B T/\hbar \omega_{m}\right)\delta (t)$)
can be safely used in (\ref{qbm}), so to get
\begin{equation}
\langle Q^{2}\rangle _{BM} = \frac{k_B T}{2\hbar \omega_{m}}
\frac{{\cal Q}^{2}}{\left(1+g_{1}\right)
\left({\cal Q}^{2}+g_{1}\right)}.
\label{qbm2}
\end{equation}
Finally it is
\begin{equation}
\langle Q^{2}\rangle _{st} = 
\frac{g_{1}^{2}}{8 \eta \zeta} 
\frac{1+{\cal Q}^{2}+g_{1}}{\left(1
+g_{1}\right)\left({\cal Q}^{2}+g_{1}\right)}+
\left[\frac{\zeta}{8}+
\frac{k_B T}{2\hbar \omega_{m}}\right]
\frac{{\cal Q}^{2}}{\left(1+g_{1}\right)
\left({\cal Q}^{2}+g_{1}\right)},
\label{q2sc2}
\end{equation}
where we have introduced the rescaled, dimensionless, input 
power of the driving laser 
\begin{equation} 
\zeta =  \frac{16 G^{2}\beta ^{2}}{\gamma_m\gamma_{c}}=
\frac{64G^{2}}{\hbar \omega_{0}\gamma_{m}\gamma_{c}^{2}}\wp .
\label{zeta}
\end{equation} 
Eq.~(\ref{q2sc2}) coincides with the corresponding one obtained in 
\cite{MVTPRL} using a Master equation description of the stochastic 
cooling feedback scheme.

An analogous procedure can be followed to get the stationary value
$\langle P^{2}\rangle _{st}$.
Using Eqs.~(\ref{PTIME}), (\ref{CHITILSC}), (\ref{KP}), (\ref{corsc}),
and (\ref{zeta}),  
one obtains the general expression
\begin{equation}
\langle P^{2}\rangle _{st} = 
\frac{g_{1}^{2}}{8 \eta \zeta}
\frac{{\cal Q}^{2}}{\left(1+g_{1}\right)
\left({\cal Q}^{2}+g_{1}\right)}+
\frac{\zeta}{8}
\frac{g_{1}^{2}+{\cal Q}^{2}+g_{1}}{\left(1
+g_{1}\right)\left({\cal Q}^{2}+g_{1}\right)} +
\langle P^{2}\rangle _{BM},
\label{p2sc}
\end{equation}
where the quantum Brownian motion contribution is now given by
\begin{equation}
\langle P^{2}\rangle _{BM} = \int_{-\varpi}^{\varpi}\frac{d \omega}{2 
\pi} \frac{\gamma_{m}}{2 \omega_{m}}\omega \coth
\left(\frac{\hbar \omega}{2 k_{B} T}\right) \left 
|{\tilde \chi}_{sc}(\omega)\right|^{2} 
\left(\frac{\omega^{2}+\gamma_{m}^{2}g_{1}^{2}}{\omega_{m}^{2}}\right).
\label{pbm}
\end{equation}
In this case, the classical, high-temperature, approximation 
$\coth(\hbar \omega/2k_B T) \simeq 2k_B T/\hbar \omega $ has to be made with 
care, because, due to the presence of the $\omega^{2}$ term, 
the integral (\ref{pbm}) has an ultraviolet divergence in the usually 
considered $\varpi \to \infty $ limit (see also Eq.~(\ref{suscsc})).
This means that, differently from $\langle Q^{2}\rangle _{BM}$,
the classical approximation for $\langle P^{2}\rangle 
_{BM}$ is valid only under the {\em stronger} condition $\hbar 
\varpi \ll k_B T $ \cite{GRAB}, 
and that in the intermediate temperature range
$ \hbar \varpi \gg k_B T \gg \hbar \omega_{m}$ (which may be of interest
for optomechanical systems), one has a correction of order 
$\ln\left(\hbar \varpi /k_B T\right)$. One has therefore \cite{GRAB}
\begin{equation}
\langle P^{2}\rangle _{BM} = \frac{k_B T}{2\hbar \omega_{m}}
\frac{g_{1}^{2}+{\cal Q}^{2}+g_{1}}{\left(1
+g_{1}\right)\left({\cal Q}^{2}+g_{1}\right)}+
\frac{\gamma_{m}}{\pi \omega_{m}}\ln\left(\frac{\hbar \varpi}{2 \pi k 
T}\right), \label{pbm2}
\end{equation}
so that one finally gets
\begin{equation}
\langle P^{2}\rangle _{st} = 
\frac{g_{1}^{2}}{8 \eta \zeta}
\frac{{\cal Q}^{2}}{\left(1+g_{1}\right)
\left({\cal Q}^{2}+g_{1}\right)}+
\left[\frac{\zeta}{8}+
\frac{k_B T}{2\hbar \omega_{m}}\right]
\frac{g_{1}^{2}+{\cal Q}^{2}+g_{1}}{\left(1
+g_{1}\right)\left({\cal Q}^{2}+g_{1}\right)} +
\frac{\gamma_{m}}{\pi \omega_{m}}\ln\left(\frac{\hbar \varpi}{2 \pi k 
T}\right).
\label{p2sc2}
\end{equation} 
This expression coincides with the corresponding one obtained in 
\cite{MVTPRL} using a Master equation description, except for the 
logarithmic correction, which however, in the case of mirror with 
a good quality factor ${\cal Q}$, is quite 
small, even in the intermediate
temperature range $ \hbar \varpi \gg k_B T \gg \hbar \omega_{m}$.

A peculiar aspect of the stochastic cooling feedback scheme, which 
has not been underlined in \cite{MVTPRL}, is its capability of 
inducing steady-state correlations between the position and the 
momentum of the mirror, i.e., the fact that $\langle QP+PQ 
\rangle_{st} \neq 0$. This correlation can be evaluated in the same 
way as above, starting from Eqs.~(\ref{QTIME}) and (\ref{PTIME}), and 
getting
\begin{equation} 
\frac{\langle QP+PQ\rangle _{st}}{2} = 
-\int_{0}^{\infty}dt' \int_{0}^{\infty}dt''
K_{Q}(t')\chi_{sc}(t'') c_{1}(t'-t'') +
\int_{0}^{\infty}dt' \int_{0}^{\infty}dt''
K_{P}(t')\chi_{sc}(t'') c_{2}(t'-t''). \label{qp1}
\end{equation}
Then, using Eqs.~(\ref{CHITILSC}), 
(\ref{KQ}), (\ref{KP}), and (\ref{corsc}), and performing the classical 
approximation on the quantum Brownian motion contribution (there is 
no ultraviolet divergence for $\varpi \to \infty$ in this case),
one gets
\begin{equation}
\frac{\langle QP+PQ\rangle _{st}}{2} =
\left(\frac{\zeta}{8}+
\frac{k_BT}{2\hbar\omega_{m}}
\right)
\frac{g_{1}{\cal Q}}{(1+g_{1})({\cal Q}^{2}+g_{1})}
-\frac{g_{1}^{2}}{8 \eta \zeta}
\frac{{\cal Q}}{(1+g_{1})
({\cal Q}^{2}+g_{1})}\,.
\label{QPSS1} 
\end{equation}
Each steady state expression (\ref{q2sc2}), (\ref{p2sc2}) and
(\ref{QPSS1}) has three contributions: the thermal term due to the 
mirror Brownian motion, the back action of the radiation pressure, 
proportional to the input power $\zeta$, and the feedback-induced noise term
proportional to $g_{1}^{2}$ and inversely proportional to the input 
power. At sufficiently large temperatures, the 
thermal noise contribution is much larger than the others and the 
mirror dynamics is faithfully described in terms of {\em classical}
stochastic equations. This classical description amounts to neglect 
all the radiation input noises into the evolution equations of the 
system, so that ${\cal W}(t)$ is the only noise acting on the system.
This classical description has been succesfully
used in Refs.~\cite{HEIPRL,PINARD} to account for the experimental 
data, in the case of a cold damping feedback scheme at room temperature.
It is however evident that the radiation back action and the 
feedback-induced noise cannot be neglected in general. 
For example, the classical approximation for 
$\langle Q^{2}\rangle _{st}$ suggests that it would be possible to 
localize the mirror without limit, i.e., 
$\langle Q^{2}\rangle _{st} \to 0$, using an ever increasing feedback gain 
$g_{1}$ and keeping the input power fixed, 
while this is no more true as soon as the feedback-induced noise 
term proportional to $g_{1}^{2}$ is included.

\begin{figure}[h]
\centerline{\epsfig{figure=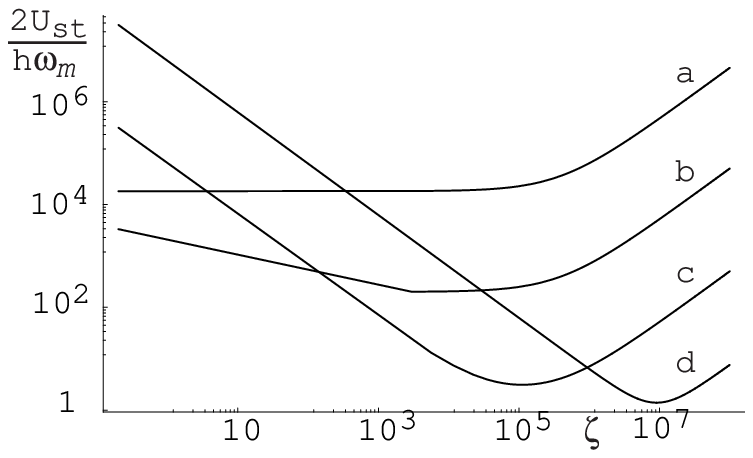,width=3.5in}}
\caption{\widetext 
Rescaled steady-state energy $2U_{st}/\hbar \omega_{m}$ 
versus the rescaled input power $\zeta$, plotted for 
different values of $g_{1}$ (a: $g_{1}=10$, b: $g_{1}=10^{3}$,
c: $g_{1}=10^{5}$, d: $g_{1}=10^{7}$) at fixed ${\cal Q}=10^{7}$, 
and with $k_{B}T/\hbar \omega_{m}= 10^{5}$ and $\eta =0.8$.
The optimal input power $\zeta_{opt}$ correspondingly increases, and
for high gain values, ground state cooling can be 
achieved.}
\label{enesc-g}
\end{figure}

\begin{figure}[t]
\centerline{\epsfig{figure=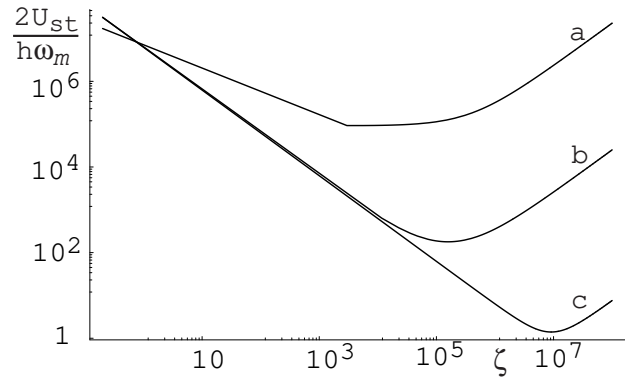,width=3.5in}}
\caption{\widetext 
Rescaled steady-state energy $2U_{st}/\hbar \omega_{m}$ 
versus $\zeta$ for 
increasing values of the mechanical quality factor ${\cal Q}$ 
(a: ${\cal Q}=10^{3}$,
b: ${\cal Q}=10^{5}$, c: ${\cal Q}=10^{7}$) at fixed $g_{1}=10^{7}$,
and with $k_{B}T/\hbar \omega_{m}= 10^{5}$ and $\eta =0.8$.}
\label{enesc-q}
\end{figure}

The stochastic cooling feedback scheme has been introduced in
\cite{MVTPRL} as a promising method for significantly cooling the 
cavity mirror. Let us therefore consider the optimal conditions for 
cooling, and the cooling limits of this scheme. 
The interesting quantity is the stationary
oscillator energy $U_{st}$, which, neglecting the logarithmic 
correction of Eq.~(\ref{p2sc2})), can be written as
\begin{equation}
U_{st} = \hbar \omega_m\left[\langle Q^{2}\rangle _{st}
+\langle P^{2}\rangle _{st}\right] = 
\frac{\hbar \omega_m}{8}\left[
\frac{g_{1}^{2}}{\eta \zeta }\frac{ 
\left(1+2{\cal Q}^{2}+g_{1}\right)}{\left(1
+g_{1}\right)\left({\cal Q}^{2}+g_{1}\right)}+
\left(\zeta+ \frac{4k_B T}{\hbar \omega_{m}}\right)\frac{
\left(g_{1}^{2}+2{\cal Q}^{2}+g_{1}\right)}{\left(1
+g_{1}\right)\left({\cal Q}^{2}+g_{1}\right)}\right].
\label{ener}
\end{equation} 
It is evident from Eq.~(\ref{ener}) that the effective temperature is
decreased only if both ${\cal Q}$ and $g_1$ are very large. At the same
time, the additional terms due to the feedback-induced noise and the 
back-action noise have to remain bounded for ${\cal Q} \to \infty$ and 
$g_1 \to \infty$, and this can be obtained by minimizing $U_{st}$ with
respect to $\zeta$ keeping ${\cal Q}$ and $g_1$ fixed 
(physically this means optimizing the input power $\wp$ at given
$g_1$ and ${\cal Q}$). It is possible to
check that these additional terms are bounded only for very large ${\cal Q}$,
that is, if ${\cal Q}/g_1 \to \infty$ and in this case the minimizing
rescaled input power is $\zeta_{opt} \simeq g_1/\sqrt{\eta}$. Under these 
conditions, the steady state oscillator energy becomes
\begin{equation}
U_{st} \simeq \frac{\hbar \omega_m}{2}\left[\frac{1}{\sqrt{\eta}}+
\frac{2k_B T}{\hbar \omega_m}\frac{1}{g_1}\right]\;,
\label{enersimpl}
\end{equation}
showing that, in the ideal limit 
$\eta =1$, $g_1 \to \infty$, $\zeta \sim g_1 \to \infty$, ${\cal Q}/g_1 \to
\infty$, the stochastic cooling feedback scheme is able to reach the
quantum limit $U_{st} =\hbar \omega_m /2$, i.e., it is able to cool
the mirror down to its quantum ground state. The behavior of the 
steady-state energy is shown in Figs.~\ref{enesc-g} and \ref{enesc-q}, 
where $U_{st}$ (in zero-point energy units $\hbar \omega_{m}/2$)
is plotted as a function of the rescaled input power $\zeta$. 
In Fig.~\ref{enesc-g}, $2U_{st}/\hbar \omega_{m}$ is plotted for 
increasing values of $g_{1}$ (a: $g_{1}=10$, b: $g_{1}=10^{3}$,
c: $g_{1}=10^{5}$, d: $g_{1}=10^{7}$) at fixed ${\cal Q}=10^{7}$, 
and with $k_{B}T/\hbar \omega_{m}= 10^{5}$ and $\eta =0.8$.
The figure shows the corresponding increase of the optimal input power 
minimizing the energy,
and that for high gain values, ground state cooling can be 
essentially achieved, even with a nonunit detection efficiency. 
In Fig.~\ref{enesc-q}, $2U_{st}/\hbar \omega_{m}$ 
is instead plotted for 
increasing values of the mechanical quality factor ${\cal Q}$ 
(a: ${\cal Q}=10^{3}$,
b: ${\cal Q}=10^{5}$, c: ${\cal Q}=10^{7}$) at fixed $g_{1}=10^{7}$.
The figure clearly shows the importance of ${\cal Q}$ in stochastic 
cooling feedback and that ground state 
cooling is achieved only when ${\cal Q}$ is sufficiently large.

The possibility to reach ground
state cooling of a macroscopic mirror using the feedback scheme of 
Ref.~\cite{MVTPRL} was first pointed out, using
an approximate treatment, in \cite{RON}, where the need of a
very large mechanical quality factor is underlined. Here we confirm this
result using the more general QLE approach. 

\begin{figure}[h]
\centerline{\epsfig{figure=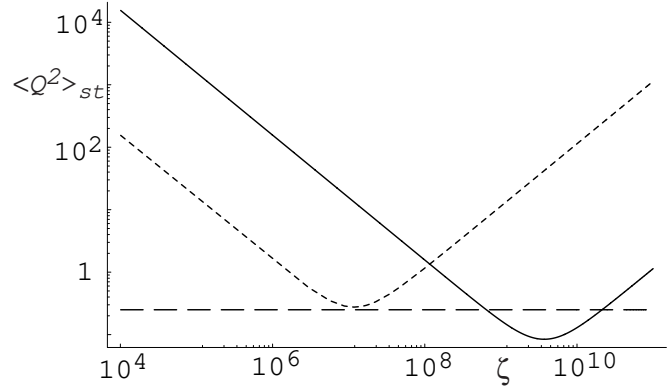,width=3.5in}}
\caption{\widetext 
Steady state position variance $\langle Q^{2}\rangle _{st}$
versus $\zeta$ for 
two values of the feedback gain, $g_1=10^{7}$ (dotted line),
and $g_1=10^{9}$ (full line). The dashed
line denotes the standard quantum limit 
$\langle Q^{2}\rangle _{st}= 1/4$, while the other parameters are: 
${\cal Q}=10^{4}$, $k_{B}T/\hbar \omega_{m}=10^{5}$ and $\eta =0.8$.}
\label{squee}
\end{figure}

The steady state of the mirror mode
in the presence of stochastic cooling feedback shows other peculiar
aspects and interesting limiting cases.
Thanks to the linearization of the problem (see Eqs.~(\ref{QLE})), 
this steady state is a Gaussian state, 
which however is never exactly a thermal state 
because it is always $\langle Q^{2}\rangle _{st} \neq 
\langle P^{2}\rangle _{st}$ and $\langle QP+PQ\rangle _{st} \neq 0$.
Its phase space contours are therefore ellipses, rotated by
an angle $\phi = (1/2)\arctan\left[\langle QP+PQ\rangle _{st}/\left(
\langle Q^{2}\rangle _{st} - \langle P^{2}\rangle _{st}\right)\right]$
with respect to the $Q$ axis. The steady state becomes approximately
a thermal state only in the limit of very large ${\cal Q}$ (and ${\cal Q}^2
\gg g_1$), as it can be seen from Eqs.~(\ref{q2sc2}), (\ref{p2sc2}) and
(\ref{QPSS1}). This thermal state approaches the quantum ground state of
the oscillating mirror when also the feedback gain and the input power
become very large.
There are however other interesting limits in which the stochastic cooling
feedback steady state shows nonclassical features.
For example, the Gaussian steady state becomes a
contractive state, which has been shown to be able
to break the standard quantum limit in \cite{YUEN}, when 
$\langle QP+PQ\rangle _{st}$ becomes negative, and this can be 
achieved at sufficiently large feedback gain, that is, when
$g_{1}>\eta \zeta \left(\zeta+ 4k_BT/\hbar\omega_{m}\right)$
(see Eq.~(\ref{QPSS1})).
Finally, stochastic cooling feedback can be used even to achieve
steady state position squeezing, that is, to beat the standard quantum
limit $\langle Q^{2}\rangle _{st} <1/4$. The strategy is similar to that
followed for cooling. First of all one has to minimize 
$\langle Q^{2}\rangle _{st}$ with respect to the input power $\zeta$
at fixed $g_1$ and ${\cal Q}$, obtaining
\begin{equation}
\langle Q^{2}\rangle _{st}^{min}=\frac{g_1 {\cal Q}\sqrt{1+{\cal Q}^2+g_1}
}{4\sqrt{\eta}(1+g_1)({\cal Q}^2+g_1)}+\frac{k_B T}{2\hbar \omega_m}
\frac{{\cal Q}^2}{(1+g_1)({\cal Q}^2+g_1)}.
\label{q2min}
\end{equation}
This quantity can become arbitrarily small in the limit of very large
feedback gain, and provided that $g_1 \gg {\cal Q}^2$. That is, differently
from cooling, position squeezing is achieved in the limit $g_1 \to \infty$
(implying $\zeta \to \infty$), and there is no condition on 
the mechanical quality
factor. Under this limiting conditions, $\langle Q^{2}\rangle _{st}$ goes to
zero as $g_1^{-1/2}$, and, at the same time, $\langle P^{2}\rangle _{st}$
diverges as $g_1^{3/2}$, so that, in this limit, the steady state 
for the stochastic cooling feedback approaches the position eigenstate
with $Q=0$, that is, the mirror tends to be perfectly localized at its
equilibrium position. The possibility to beat the standard quantum limit
for the position uncertainty is shown in Fig.~\ref{squee}, where
$\langle Q^{2}\rangle _{st}$ is plotted versus $\zeta$ for two different
values of the feedback gain, $g_1=10^7$ (dotted line), and $g_1=10^9$ (full 
line), with ${\cal Q}=10^4$, $k_B T/\hbar \omega_m= 10^5$, and 
$\eta =0.8$. For the higher value of the feedback gain, the standard quantum
limit $\langle Q^{2}\rangle _{st}= 1/4$ (dashed line)
is beaten in a range of values of the input power $\zeta$.

\subsection{Cold damping feedback}

Now we characterize the stationary state of the mirror in the 
presence of cold damping. This stationary state has been already studied
using classical arguments in \cite{HEIPRL,PINARD}, while the discussion of
the cooling limits of cold damping in the quantum case has been 
recently presented in \cite{COURTY}. Here we shall generalize the results
of \cite{COURTY} to the case of nonideal quantum efficiency $\eta < 1$,
and we shall compare the cooling capabilities of the two feedback schemes.

Using the solution (\ref{QT}) for the time evolution, one has
\begin{equation} 
\langle Q^{2}\rangle _{st} = \lim_{t \to \infty}
\langle Q(t)^{2}\rangle = 
\int_{0}^{\infty}dt' \int_{0}^{\infty}dt''
\chi_{cd}(t')\chi_{cd}(t'') c(t'-t''), \label{q21cd}
\end{equation}
where $c(t)$ is the stationary symmetrized
correlation function of the noise 
term $n(t)=\frac{2G\beta }{\sqrt{\gamma_{c}}}
  X_{in}(t') + {\cal W}(t')
  -\frac{g_{cd}}{\sqrt{\gamma_{c}}}\dot{Y}_{in}(t')
 + \frac{g_{cd}}{2\sqrt{\gamma_{c}\eta}}\dot{Y}_{in}^{\eta}(t')$
appearing in Eq.~(\ref{QT}). Using the correlation functions 
(\ref{INCOR}), (\ref{BROWCOR}), (\ref{INCORETA}), and (\ref{dotcorre}),
one gets
\begin{equation}\label{corcd}
c(t) = \frac{4 G^{2}\beta ^{2}}{\gamma_{c}} \delta (t) 
- \frac{g_{cd}^2}{4 \eta \gamma_c}\ddot{\delta}(t)+
\frac{\gamma_{m}}{2 \pi \omega _{m}}{\cal F}_{r}(t). 
\end{equation}
Since in the cold damping case it is $P(t)=\dot{Q}(t)/\omega_m$,
it is straightforward to derive from Eq.~(\ref{q21cd}) the expressions
for $\langle P^{2}\rangle _{st}$ and $\langle PQ+QP\rangle _{st}$,
which are given by
\begin{eqnarray}
\langle PQ+QP\rangle _{st}&=&\frac{1}{\omega_m}\lim_{t \to \infty}
\frac{d}{dt}\langle Q(t)^{2}\rangle =0, \\
\langle P^{2}\rangle _{st}&=&
\frac{1}{\omega_m^2}\int_{0}^{\infty}dt' \int_{0}^{\infty}dt''
\dot{\chi}_{cd}(t')\dot{\chi}_{cd}(t'') c(t'-t''). \label{p21cd}
\end{eqnarray}
These stationary expressions can be rewritten in terms of the Fourier
transforms of the noise correlation functions, in the same way as we have
done for the Brownian motion term in the preceding subsection. Using
Eqs.~(\ref{BROWCOR}), (\ref{CHITILCD}), (\ref{zeta}), and (\ref{corcd}), 
one has
\begin{eqnarray}
\langle Q^{2}\rangle _{st} &=& \gamma_{m}
\int_{-\infty}^{\infty}\frac{d \omega}{2 
\pi} \left 
|{\tilde \chi}_{cd}(\omega)\right|^{2}\left[\frac{\zeta}{4}
+\frac{g_2^2}{4 \eta \zeta }
\frac{\omega^2}{\omega_m^2}
\Theta_{\Delta \omega}(\omega)+\frac{\omega}{2 \omega_{m}} \coth
\left(\frac{\hbar \omega}{2 k_{B} T}\right)\Theta_{[-\varpi,\varpi]}(\omega) 
\right],
\label{q2fou} \\
\langle P^{2}\rangle _{st} &=& \gamma_{m}
\int_{-\infty}^{\infty}\frac{d \omega}{2 
\pi} \frac{\omega^2}{\omega_m^2}\left 
|{\tilde \chi}_{cd}(\omega)\right|^{2}\left[\frac{\zeta}{4}
+\frac{g_2^2}{4 \eta \zeta}\frac{\omega^2}{\omega_m^2}
\Theta_{\Delta \omega}(\omega)+\frac{\omega}{2 \omega_{m}} \coth
\left(\frac{\hbar \omega}{2 k_{B} T}\right)\Theta_{[-\varpi,\varpi]}(\omega)
\right] ,
\label{p2fou} 
\end{eqnarray}
where 
\begin{equation}
{\tilde \chi}_{cd}(\omega)=\frac{\omega_{m}}{\omega_{m}^{2}
	-\omega^{2}+i\omega \gamma_{m}\left(1+g_{2}\right)}
	\label{susccd}
\end{equation}
is the frequency-dependent susceptibility of the mirror in the 
cold damping feedback scheme, and $\Theta_{I}(\omega)$ is a ``gate'' function, 
equal to one within the interval $I$ and equal to zero outside.
Notice that we have introduced not only
the gate function $\Theta_{[-\varpi,\varpi]}(\omega)$ for the
thermal noise term, but also the gate function $\Theta_{\Delta \omega}(\omega)$
for the feedback-induced noise term. In fact, it is easy to see
that a frequency cutoff for the feedback 
is needed to avoid an ultraviolet divergence
in the expression for $\langle P^{2}\rangle _{st}$. Moreover,
from an experimental point of view, any feedback loop is active only
within a finite bandwidth, which in this case is given by $\Delta \omega$. 

We first evaluate $\langle Q^{2}\rangle _{st} $. The contribution of the 
feedback-induced term generally depends upon the value of the feedback 
bandwidth $\Delta \omega$. There are two relevant experimental situations: 
a narrow bandwidth containing the mechanical resonance peak, that is,
$ \gamma_m(1+g_2 ) < \Delta \omega < \omega_m $ (configuration used
in Ref.~\cite{HEIPRL,PINARD}), or a wide bandwidth with a 
very large high frequency cutoff $\varpi_{fb} \gg \omega_m, 
\gamma_m(1+g_2)$.
However, since the factor $|{\tilde \chi}_{cd}(\omega)|^2$ in Eq.~(\ref{q2fou})
is highly peaked around the resonance frequency $\omega_m$, 
$\langle Q^{2}\rangle _{st}$ is practically independent of the
feedback loop bandwidth, as soon as $\gamma_m(1+g_2) < \Delta \omega$. In fact,
either in the narrow bandwidth case, when the spectrum can be approximated
by the constant term $g_2^2/4 \eta \zeta $,
or in the case of a very large cutoff frequency, when the $\omega^2$ dependence
is kept, one gets the same result for the feedback-induced contribution, 
because
\begin{equation}
\int_{-\infty}^{\infty}\frac{d \omega}{2 
\pi} \frac{\omega^2}{\omega_m^2}\left 
|{\tilde \chi}_{cd}(\omega)\right|^{2} = \int_{-\infty}^{\infty}\frac{d \omega}{2 
\pi} \left 
|{\tilde \chi}_{cd}(\omega)\right|^{2} = \frac{1}{2\gamma_m(1+g_2)}.
\label{into} 
\end{equation}
For the Brownian motion contribution we have the same situation described
in the stochastic cooling case: the exact expression is cumbersome 
\cite{GRAB}, but
in the commonly met condition $\hbar \omega_m \ll k_B T$, the classical
approximation $\coth(\hbar \omega /2k_B T) \simeq 2k_B T/\hbar \omega$
can be made, and using Eq.~(\ref{into}) for both the thermal and the
back-action contribution, one finally gets
\begin{equation}
\langle Q^{2}\rangle _{st} = 
\left[\frac{g_{2}^{2}}{8 \eta \zeta } +
\frac{\zeta }{8}+
\frac{k_B T}{2\hbar \omega_{m}}\right]
\frac{1}{1+g_{2}}.
\label{q2cd}
\end{equation}
Notice that the corresponding expression for the stochastic cooling 
feedback (\ref{q2sc2}) coincides with Eq.~(\ref{q2cd}) in the limit
${\cal Q} \gg 1, g_1$.

Differently from $\langle Q^{2}\rangle _{st}$, 
$\langle P^{2}\rangle _{st}$ depends upon the feedback loop bandwidth.
In fact, in the large bandwidth case,
the integrand in Eq.~(\ref{p2fou}) tends to a constant at large
frequencies, and in the limit of a very large cutoff frequency $\varpi_{fb}$,
the feedback-induced contribution becomes
\begin{equation}
\langle P^{2}\rangle _{st}^{fb}=
\frac{\gamma_{m}g_{2}^{2}}{8 \eta \zeta}
\frac{\varpi_{fb}}{\pi \omega_m^2}.
\end{equation}
In the narrow bandwidth case instead, approximating the noise spectrum
with the constant term $g_2^2/4 \eta \zeta $, and using
again Eq.~(\ref{into}) within Eq.~(\ref{p2fou}), one gets a
feedback-induced noise term contribution identical to that of 
$\langle Q^{2}\rangle _{st}$ of Eq.~(\ref{q2cd}), which is 
independent of the feedback bandwidth.

A potential ultraviolet divergence and a dependence upon the
frequency cutoff $\varpi$ is present also in the quantum Brownian motion
term. In fact, as we have seen in the preceding subsection, the classical
expression for the thermal contribution to
$\langle P^{2}\rangle _{st}$, holds only in the limit of very large 
temperatures, $k_B T \gg \hbar \varpi$, while, in the intermediate temperature
regime $\hbar \omega_m \ll k_B T \ll \hbar \varpi$, one has an additional
logarithmic correction, so to get
\begin{equation}
\langle P^{2}\rangle _{st}^{BM}=
\frac{k_B T}{2\hbar \omega_{m}}
\frac{1}{1+g_{2}}+
\frac{\gamma_{m}}{\pi \omega_{m}}\ln\left(\frac{\hbar \varpi}{2 \pi k 
T}\right). \label{pbm3}
\end{equation}
Finally, the back-action term is simply evaluated using 
Eq.~(\ref{into}) and one gets the same contribution as in 
Eq.~(\ref{q2cd}),
\begin{equation}
\langle P^{2}\rangle _{st}^{ba}=
\frac{\zeta}{8(1+g_{2})}. \label{pba}
\end{equation}
Therefore, the general expression for $\langle P^{2}\rangle _{st}$
depends on the parameter regime considered and it may generally
depend upon the feedback loop high frequency cutoff $\varpi_{fb}$
and the thermal bath cutoff $\varpi$. However, in the common experimental
situation of a narrow bandwidth around the resonance peak, 
$\gamma_m(1+g_2) < 
\Delta \omega < \omega_m$, and a high ${\cal Q}$ mechanical mode 
so that the logarithmic correction in Eq.~(\ref{pbm3}) can be neglected, the
dependence on the frequency cutoffs vanishes and one has
$\langle P^{2}\rangle _{st} =\langle Q^{2}\rangle _{st}$.
Therefore, under these conditions, since it is also
$\langle QP+PQ\rangle _{st}=0$, the stationary state in the presence
of the cold damping feedback scheme is an effective thermal state
with a mean excitation number $\langle n \rangle = 2 
\langle Q^{2}\rangle _{st} -1/2$, where $\langle Q^{2}\rangle _{st}$ is given
by Eq.~(\ref{q2cd}). This effective thermal equilibrium state
in the presence of cold damping has been already pointed out in 
\cite{HEIPRL,PINARD}, within a classical treatment neglecting both the 
back-action and the feedback-induced terms. The present fully quantum analysis
shows that cold damping has two opposite effects on the effective equilibrium
temperature of the mechanical mode: on one hand $T$ is reduced by the factor
$(1 +g_2)^{-1}$, but, on the other hand, the effective temperature
is increased by the additional noise terms.

\begin{figure}[t]
\centerline{\epsfig{figure=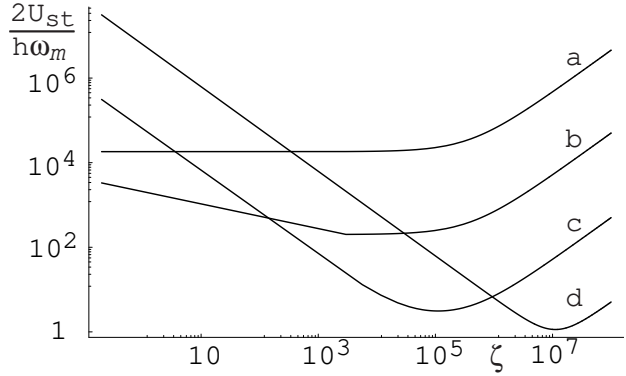,width=3.5in}}
\caption{\widetext 
Rescaled steady-state energy $2U_{st}/\hbar \omega_{m}$ 
versus the rescaled input power $\zeta$, plotted for 
different values of $g_{2}$ (a: $g_{2}=10$, b: $g_{2}=10^{3}$,
c: $g_{2}=10^{5}$, d: $g_{2}=10^{7}$),
with $k_{B}T/\hbar \omega_{m}=10^{5}$ and $\eta =0.8$.
The optimal input power correspondingly increases, and
for high gain values, ground state cooling can be 
achieved.}
\label{enecd-g}
\end{figure}

Let us now consider the optimal conditions for 
cooling and the cooling limits of the cold damping feedback scheme.
In the narrow feedback loop bandwidth case, and neglecting
the logarithmic correction to $\langle P^{2}\rangle _{st}^{BM}$,
the stationary oscillator energy is given by
\begin{equation}
U_{st} = 2\hbar \omega_m \langle Q^{2}\rangle _{st} = 
\frac{\hbar \omega_m}{4\left(1
+g_{2}\right)}\left[
\frac{g_{2}^{2}}{\eta \zeta }+
\zeta+ \frac{4k_B T}{\hbar \omega_{m}}\right].
\label{ener2}
\end{equation} 
This expression coincides with that derived and discussed
in \cite{COURTY}, except for the
presence of the homodyne detection efficiency $\eta$, which was
ideally assumed equal to one in \cite{COURTY}. The optimal conditions
for cooling can be derived in the same way as it has been done in
\cite{COURTY}. The energy $U_{st}$ is minimized with
respect to $\zeta$ keeping $g_2$ fixed, 
thereby getting $\zeta_{opt} = g_2/\sqrt{\eta}$. Under these 
conditions, the stationary oscillator energy becomes
\begin{equation}
U_{st} = \frac{\hbar \omega_m}{2}\frac{g_2}{1+g_2}
\left[\frac{1}{\sqrt{\eta}}+
\frac{2k_B T}{\hbar \omega_m}\frac{1}{g_2}\right]\;,
\label{enersimpl2}
\end{equation}
showing that, in the ideal limit 
$\eta =1$, $g_2 \to \infty$ (and therefore $\zeta \sim g_2 \to \infty$), 
also the cold damping scheme is able to reach the
quantum limit $U_{st} =\hbar \omega_m /2$, i.e., it is able to cool
the mirror to its quantum ground state, as first pointed out
in \cite{COURTY}. However, differently from the stochastic cooling 
case of the preceding subsection, the stationary energy
does not depend on the mechanical quality factor,
implying that cooling is easier to achieve using cold damping, because the
additional condition ${\cal Q}/g_2 \to \infty$ is not necessary in this case.
However, cold damping, 
at variance with stochastic cooling feedback,
does not yield any nonclassical feature in the steady state.
Fig.~\ref{enecd-g} shows the rescaled steady-state energy
$2U_{st}/\hbar \omega_{m}$ versus $\zeta$ plotted for 
increasing values of $g_{2}$ (a: $g_{2}=10$, b: $g_{2}=10^{3}$,
c: $g_{2}=10^{5}$, d: $g_{2}=10^{7}$),
with $k_{B}T/\hbar \omega_{m}= 10^{5}$ and $\eta =0.8$.
The figure is essentially indistinguishable from 
Fig.~\ref{enesc-g}, since, as 
we have seen, the steady states for the two feedback schemes becomes
identical for large mechanical quality factors.
For high gain values, ground state cooling can be 
achieved also in this case, even with nonunit homodyne detection 
efficiency. 

The ultimate quantum limit of ground state cooling is achieved in 
both schemes only if {\em both} the input power and the feedback gain 
go to infinity. If instead the input power is kept fixed, the 
effective temperature does not monotonically decrease for increasing 
feedback gain, but, as it can be easily seen from Eqs.~(\ref{ener})
and (\ref{ener2}), there is an optimal feedback gain, giving a minimum steady 
state energy, generally much greater than the quantum ground state 
energy. The existence of an optimal feedback gain at fixed input 
power is a consequence of the feedback-induced noise term originating 
from the quantum input noise of the radiation. In a classical treatment 
neglecting all quantum radiation noises, one would have instead erroneously
concluded that the oscillator energy can be made arbitrarily small, by 
increasing the feedback gain, and independently of the radiation input 
power. This is another example of the importance of including the radiation 
quantum noises, showing again that a full quantum treatment is necessary
to get an exhaustive description of the system dynamics \cite{LETTER}.

The experimental achievement of ground state cooling via feedback is 
prohibitive with present day technology. For example, the experiments 
of Refs.~\cite{HEIPRL,PINARD} have used feedback gains up to 
$g_{2}=40$ and an input power corresponding to $\zeta \simeq 1$, and 
it is certainly difficult to realize in practice the limit of very 
large gains and input powers. This is not 
surprising, since this would imply the preparation of a mechanical
macroscopic degree of freedom in its quantum ground state, which 
is remarkable. The same considerations hold 
for breaking the standard quantum limit
for the steady state position fluctuations with the stochastic cooling 
feedback.

\section{Spectral measurements and their sensitivity}

Both stochastic cooling and cold 
damping feedback schemes cool the mirror by overdamping it, 
thereby strongly decreasing its mechanical susceptibility at resonance
(see Eqs.~(\ref{suscsc}) and (\ref{susccd})).
As a consequence, the oscillator does not resonantly respond to the 
thermal noise, yielding
in this way an almost complete suppression of the resonance peak in the 
noise power spectrum. Since the effective temperature is proportional
to the area below the noise power spectrum, this implies cooling.
However, the strong reduction of the mechanical susceptibility at resonance 
means that the mirror does not respond not only to the noise but 
also to any force acting on it. 
Therefore one expects that the SNR
of the optomechanical device is not improved by feedback.
However, we shall see that this intuitive guess is valid only under
{\em stationary} conditions, and that, at least in the case of an 
{\em impulsive} force, a {\em nonstationary} strategy can be designed
to improve the sensitivity for the detection of a weak classical force.
The possibility to use the above feedback cooling schemes in a 
nonstationary way has been first shown in \cite{LETTER}. Here 
we shall reconsider and extend the treatment of \cite{LETTER}, 
adopting a general description of nonstationary spectral 
measurements.

Spectral measurements are performed whenever the classical force 
$f(t)$ to detect has a characteristic frequency. Since the directly 
measured quantity is the output homodyne photocurrent $Y_{out}(t)$,
we define the {\em signal} $S(\omega)$ as
\begin{equation}
  	S(\omega)= \left|\int_{-\infty}^{+\infty}dt e^{-i\omega t}\langle 
  	Y_{out}(t)\rangle F_{T_{m}}(t)\right|,
  	\label{signal}
  \end{equation}
where $F_{T_{m}}(t)$ is a ``filter'' function, approximately 
equal to one in the time interval $[0,T_{m}]$ in which the spectral 
measurement is performed, and equal to zero otherwise. Using 
Eq.~(\ref{adiab}), 
the input-output relation (\ref{BOUNDARY}), and
the time evolution of the position operator $Q(t)$
(Eq.~(\ref{QTIME}) or (\ref{QT})), the signal can be rewritten as
\begin{equation}
  	S(\omega)= \frac{8G\beta \eta}{2\pi \sqrt{\gamma_{c}}}
  	\left|\int_{-\infty}^{+\infty}d\omega' {\tilde \chi}(\omega')
  	\tilde{f}(\omega') \tilde{F}_{T_{m}}(\omega -\omega')\right|,
  	\label{signal2}
  \end{equation} 
where $\tilde{f}(\omega)$ and $\tilde{F}_{T_{m}}(\omega )$ 
are the Fourier transforms of the force and 
of the filter function, respectively, and ${\tilde \chi}(\omega)$ is equal to
${\tilde \chi}_{sc}(\omega)$ or ${\tilde \chi}_{cd}(\omega)$, according to the 
feedback scheme considered. 

The noise corresponding to the signal $S(\omega)$ will be given by 
its ``variance''; since the signal is zero when $f(t)=0$, the noise
spectrum can be generally written as 
\begin{equation}
  	 N(\omega)= \left\{\int_{-\infty}^{+\infty} dt F_{T_{m}}(t) 
  	\int_{-\infty}^{+\infty} dt' F_{T_{m}}(t') e^{-i\omega (t-t')} 
  	 \langle Y_{out}(t)Y_{out}(t')\rangle_{f=0} \right\}^{1/2},
  	\label{noise}
  \end{equation}
where the subscript $f=0$ means evaluation
in the absence of the external force. Using again 
(\ref{adiab}), Eqs.~(\ref{BOUNDARY}), 
and the input noises correlation functions (\ref{INCOR}) and 
(\ref{INCORETA}), the 
spectral noise can be rewritten as
\begin{equation}
 N(\omega)= \left\{ \frac{(8G\beta \eta)^{2}}{\gamma_{c}}
  	\int_{-\infty}^{+\infty}dt F_{T_{m}}(t) 
  	\int_{-\infty}^{+\infty}dt' F_{T_{m}}(t') e^{-i\omega 
  	(t-t')}C(t,t')+\eta \int_{-\infty}^{+\infty}dt F_{T_{m}}(t)^{2} 
  	\right\}^{1/2},
  	\label{noise2}
  \end{equation}
where $C(t,t')=\langle Q(t)Q(t')+Q(t')Q(t)\rangle/2 $ is the 
symmetrized correlation function of the oscillator position.
This very general expression of the noise spectrum is nonstationary
because it depends upon the nonstationary correlation function 
$C(t,t')$. The last term in Eq.~(\ref{noise2})
is the shot noise term due to the radiation input noise.

\subsection{Stationary spectral measurements}

Spectral measurements are usually performed in the stationary case, 
that is, using a measurement time $T_{m}$ much 
larger than the typical oscillator timescales. The most significant
timescale is the mechanical
relaxation time, which is $\gamma_{m}^{-1}$ in the absence of feedback 
and $[\gamma_{m}(1+g_{i})]^{-1}$ ($i=1,2$) in the presence of 
feedback. In the stationary case,
the oscillator is relaxed to equilibrium and, redefining $t'=t+\tau$,
the correlation function $C(t,t')=C(t,t+\tau)$ in Eq.~(\ref{noise2}) 
is replaced by the {\em stationary} 
correlation function $C_{st}(\tau)=\lim_{t\to \infty} C(t,t+\tau)$. 
Moreover, for very large $T_m$,
one has $F_{T_{m}}(t+\tau) \simeq F_{T_{m}}(t)\simeq 1$ 
and, defining the 
measurement time $T_{m}$ so that $T_{m}=\int dt F_{T_{m}}(t)^{2}$,  
Eq.~(\ref{noise2}) assumes the form
\begin{equation}
  	N(\omega)= \left\{ \left[\frac{(8G\beta \eta)^{2}}{\gamma_{c}}
  	N_{Q}^{2}(\omega)+\eta\right]T_{m}  
  	\right\}^{1/2},
  	\label{noisesta}
  \end{equation} 
where 
\begin{equation}
N_{Q}^{2}(\omega)=\int_{-\infty}^{+\infty} d\tau e^{-i\omega \tau}C(\tau) ,
\label{noiseq}
\end{equation}
is the stationary position noise spectrum. 
This noise spectrum can be easily evaluated using the results of the
preceding section. In fact, using the definition of $C_{st}(\tau)$
and the inverse Fourier transform of Eq.~(\ref{noiseq}), one has
\begin{equation}
\langle Q^2 \rangle_{st} = \lim_{t \to \infty}\langle Q^2(t) \rangle
= C_{st}(0) = \int_{-\infty}^{+\infty} \frac{d \omega }{2 \pi}N_Q^{2}(\omega).
\label{perspe}
\end{equation}
The position noise spectrum can then be extracted from the stationary
mean values derived in the preceding section. Using Eq.~(\ref{q2fou}),
one has
\begin{equation}
N_{Q,cd}^{2}(\omega)=\gamma_{m}\left 
|{\tilde \chi}_{cd}(\omega)\right|^{2}\left[\frac{\zeta}{4}
+\frac{g_2^2}{4 \eta \zeta }
\frac{\omega^2}{\omega_m^2}
\Theta_{\Delta \omega}(\omega)+\frac{\omega}{2 \omega_{m}} \coth
\left(\frac{\hbar \omega}{2 k_{B} T}\right)\Theta_{[-\varpi,\varpi]}(\omega) 
\right],
\label{qspecd}
\end{equation}
for the cold damping scheme, while the derivation for the stochastic cooling
case is less immediate. In fact, using Eqs.~(\ref{q21}), (\ref{corsc}), and
(\ref{qbm}), one gets
\begin{equation}
\langle Q^2 \rangle_{st}=\int_{-\infty}^{+\infty}
\frac{d \omega }{2 \pi} \gamma_{m} \left [
\left|{\tilde \chi}_{sc}(\omega)\right|^{2}\left(\frac{\zeta }{4}
+\frac{\omega}{2 \omega_{m}} \coth
\left(\frac{\hbar \omega}{2 k_{B} T}\right)\Theta_{[-\varpi,\varpi]}(\omega) 
\right) +
\left|{\tilde K}_Q(\omega)\right|^{2}
\frac{g_2^2}{4 \eta \zeta }
\Theta_{\Delta \omega}(\omega)\right].
\label{qspesc1}
\end{equation}
Then, using the Fourier transform of Eq.~(\ref{KQ}) in Eq.~(\ref{qspesc1}),
one finally gets
\begin{equation}
N_{Q,sc}^{2}(\omega)=\gamma_{m} \left 
|{\tilde \chi}_{sc}(\omega)\right|^{2}\left[\frac{\zeta }{4}
+\frac{g_2^2}{4 \eta \zeta }
\frac{\omega^2+\gamma_m^2}{\omega_m^2}
\Theta_{\Delta \omega}(\omega)+\frac{\omega}{2 \omega_{m}} \coth
\left(\frac{\hbar \omega}{2 k_{B} T}\right)\Theta_{[-\varpi,\varpi]}(\omega) 
\right].
\label{qspesc}
\end{equation}
This position noise spectrum for the stochastic cooling feedback essentially
coincides with that already obtained in \cite{MVTPRL}, 
except that in that paper the high temperature limit ($\coth(\hbar \omega
/2k_B T) \simeq 2 k_B T/\hbar \omega$) is considered and the presence
of the frequency cutoffs $\varpi$ and $\varpi_{fb}$ is not taken into
account.
The noise spectrum in the cold damping case of Eq.~(\ref{qspecd})
instead essentially reproduces the one obtained in \cite{COURTY}, with the
difference that in Ref.~\cite{COURTY} the homodyne detection
efficiency $\eta$ is set equal to one, and the feedback and thermal noise
cutoff functions have not been explicitely considered. The comparison
between Eqs.~(\ref{qspecd}) and (\ref{qspesc}) shows once again the
similarities of the two schemes. The only differences lie in the 
different susceptibilities and in the feedback-induced noise term,
which has an additional $\gamma_m^2/\omega_m^2$ factor in the stochastic
cooling case, which is however usually negligible with good mechanical quality
factors. In fact, it is possible to see that the two noise spectra 
are practically indistinguishable in a very large parameter region.

The effectively detected position noise spectrum is not given
by Eqs.~(\ref{qspecd}) and (\ref{qspesc}), but one has to add the 
shot noise contribution due to the input noise in the homodyne photocurrent. 
In fact, using Eq.~(\ref{noisesta}), and rescaling it to a position spectrum, 
one has
\begin{equation}
N_{Q,det}^{2}(\omega)=\gamma_{m}\left 
|{\tilde \chi}_{i}(\omega)\right|^{2}\left[\frac{\zeta}{4}
+\frac{g_i^2}{4 \eta \zeta}
\frac{\omega^2+\delta_{i,1}\gamma_m^2}{\omega_m^2}
+\frac{\omega}{2 \omega_{m}} \coth
\left(\frac{\hbar \omega}{2 k_{B} T}\right)
\right]+\frac{1}{4 \eta \zeta \gamma_{m}},
\label{qspedet}
\end{equation}
where $i=1$ refers to the stochastic cooling case and 
$i=2$ to the cold damping case, 
The homodyne-detected position noise spectrum is actually subject 
also to cavity filtering, yielding an experimental high frequency 
cutoff $\gamma_{c}$, which however does not appear in 
Eq.~(\ref{qspedet}) because we have adiabatically eliminated the 
cavity mode from the beginning. Therefore the spectrum of 
Eq.~(\ref{qspedet}) provides a faithful description
of the mirror mode dynamics only for $\omega < \gamma_{c}$; 
since it is usually  $\varpi, \varpi_{fb} > \gamma_{c}$,
we have not considered the feedback and reservoir 
cutoff functions in Eq.~(\ref{qspedet}), and we shall not consider 
them in the following. 
The detected noise spectrum has three contributions: the Brownian
motion term which is independent of the input power $\wp$, the shot noise
term inversely proportional to $\wp$, and the back-action 
term, proportional to $\wp$. The main effect of feedback 
on the spectrum is the
modification of the susceptibility due to the increase of damping,
which is responsible for the suppression and widening of the resonance peak.
This peak suppression in the noise spectrum
has been already predicted and illustrated in
\cite{MVTPRL,COURTY}, and experimentally verified for the cold damping
case in \cite{HEIPRL,PINARD}.
Moreover, the feedback-induced noise term proportional to 
$g_{i}^{2}$ is responsible for an increase of the shot noise 
contribution to the spectrum. For a given feedback gain and frequency, 
the minimum noise is obtained at an intermediate, optimal, power, given 
by
\begin{equation}
\zeta_{opt}=\sqrt{\frac{1+{\cal Q}^{-2}g_{i}^{2}|{\tilde \chi}_{i}(\omega)|^{2}
\left(\omega^{2}+\delta_{i,1}\gamma_{m}^{2}\right)}{\eta \gamma_{m}^{2}
|{\tilde \chi}_{i}(\omega)|^{2}}},
\label{psql}
\end{equation}
and the corresponding value of the minimum displacement noise is
\begin{equation}
N_{Q,min}^{2}(\omega)=\gamma_{m}\left 
|{\tilde \chi}_{i}(\omega)\right|^{2}
\frac{\omega}{2 \omega_{m}} \coth
\left(\frac{\hbar \omega}{2 k_{B} T}\right)
+\frac{\left |{\tilde \chi}_{i}(\omega)\right|}{2\sqrt{\eta}}
\sqrt{1+{\cal Q}^{-2}g_{i}^{2}|{\tilde \chi}_{i}(\omega)|^{2}
\left(\omega^{2}+\delta_{i,1}\gamma_{m}^{2}\right)}.
\label{spesql}
\end{equation} 
This expression shows that both feedback schemes are able to 
arbitrarily reduce the displacement noise at resonance. In 
fact, using the fact that ${\tilde \chi}_{i}(\omega_{m}) 
\propto g_{i}^{-1}$ in both 
cases, one has that $N_{Q,min}^{2}(\omega_{m})$ can be made 
arbitrarily small by increasing the feedback gain. 
This noise reduction at resonance is similar to that
occurring to an oscillator with increasing damping, except that in 
our case, also
the feedback-induced noise increases with the gain, and it
can be kept small only if the input power is correspondingly increased
in order to maintain the optimal condition 
(\ref{psql}). This arbitrary reduction of the position noise
in a given frequency bandwidth with increasing feedback gain does not 
hold if the input power $\zeta$ is kept fixed. In this latter
case, the noise has a 
frequency-dependent lower bound which cannot be overcome by 
increasing the gain.
There is an important difference between the two feedback 
schemes. In fact, it is easy to check from Eq.~(\ref{spesql})
that in the cold damping case noise reduction takes place only close 
to resonance, and that the noise spectrum is not affected at lower 
frequencies (for example $N_{Q,min}^{2}(\omega=0)$ is not changed by the 
cold damping feedback). In the stochastic cooling case instead, 
frequency renormalization $\omega_{m}^{2} \to 
\omega_{m}^{2}+g_{1}\gamma_{m}^{2}$ allows to reduce position noise 
even at low frequencies. This reduction of position noise out of 
resonance, without cold damping but with a feedback-induced increase of the 
mechanical frequency, has been demonstrated experimentally by Cohadon {\it et 
al.} in Ref.~\cite{HEIPRL}.

\begin{figure}[t]
\centerline{\epsfig{figure=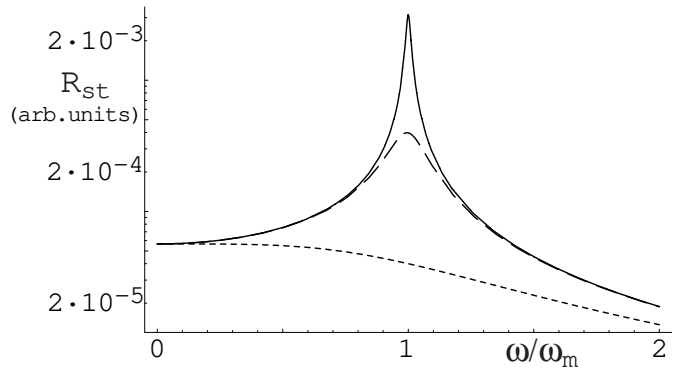,width=3.5in}}
\caption{\widetext 
Stationary SNR as a function of frequency in the case of an ideal 
impulsive force, i.e., ${\tilde f}(\omega)=$ const. The full line 
refers to the case with no feedback, the dashed line to the case 
with $g_{1}=g_{2}=10^{4}$, and the dotted line to the case with
$g_{1}=g_{2}=10^{5}$ (the two feedback schemes give 
indistinguishable results in these cases). The other parameters are 
${\cal Q}=10^{5}$, $\zeta =10$, $k_{B}T/\hbar \omega_{m}=
10^{5}$, and $\eta =0.8$. At a given frequency, the stationary 
SNR decreases for increasing feedback gain.}
\label{snrsta}
\end{figure}

In the case of stationary spectral measurements also the
expression of the signal simplifies. In fact, one has
$ \tilde{F}_{T_{m}}(\omega) \simeq \delta(\omega)$,
and Eq.~(\ref{signal2}) assumes the traditional form
\begin{equation}
  	S(\omega)= \frac{8G\beta \eta}{2\pi \sqrt{\gamma_{c}}}
  	\left| {\tilde \chi}(\omega)
  	\tilde{f}(\omega) \right|.
  	\label{signal3}
  \end{equation} 
The stationary SNR, ${\cal R}_{st}(\omega)$, is now simply obtained 
dividing the signal of Eq.~(\ref{signal3}) by the noise of 
Eq.~(\ref{noisesta}), 
\begin{equation}
 {\cal R}_{st}(\omega)= 
  	|\tilde{f}(\omega)|\left\{\gamma_{m}T_{m}\left[\frac{
  	\omega}{2\omega_{m}}\coth\left(\frac{\hbar \omega}{2k_{B}T}\right)+
  	\frac{\zeta}{4} 
   +\frac{1}{4 \eta \zeta }\left(\frac{g_{i}^{2}}{\omega_{m}^{2}} 
   \left(\omega^{2}+\delta_{i,1}\gamma_{m}^{2}\right)
  +\frac{1}{\gamma_{m}^{2}\left|{\tilde \chi}_{i}(\omega)\right|^{2}}\right)
  \right]\right\}^{-1/2},
  	\label{snr}
  \end{equation}  
where again $i=1$ refers to the stochastic cooling case and $i=2$ to 
the cold damping case.
It is easy to see that, in both cases,
feedback {\em always lowers} the stationary SNR at any frequency,
(except at $\omega=0$, where the SNR for the cold damping case does not 
depend upon the feedback gain). This is shown in Fig.~\ref{snrsta}, 
where the stationary SNR in the case of an ideal impulsive force (that 
is, ${\tilde f}(\omega)$ is a constant) is plotted for three values of 
the feedback gain. The curves refer to both feedback schemes because
the two cases $i=1,2$ gives always practically indistinguishable results, 
except for very low values of ${\cal Q}$.
As mentioned at the beginning of the section, 
this result is not surprising because the main effect of 
feedback is to decrease 
the mechanical susceptibility at resonance, so that the oscillator is 
less sensitive not only to the noise but also to the signal.
Therefore, even though the two feedback schemes are able to provide 
efficient cooling and noise reduction in narrow bandwidths for
the mechanical mode, they cannot be used to improve the sensitivity
of the optomechanical device for stationary measurements. In the next 
section we shall see how cooling via feedback can be used to improve
the sensitivity for the detection of impulsive forces, using an 
appropriate nonstationary strategy.

\section{High-sensitive nonstationary measurements}

The two feedback schemes discussed here achieve noise reduction 
through a modification of the mechanical susceptibility. However, this
modification does not translate into a sensitivity improvement because at the
same time it strongly degrades the detection of the signal. 
The sensitivity of position measurements would be improved if 
the oscillator mode could keep its intrinsic susceptibility, unmodified by
feedback, together with the reduced noise achieved by the feedback loop.
This is obviously impossible in stationary conditions, but a
situation very similar to this ideal one can be realized in the case of
the detection of an {\em impulsive} force, that is, with a 
time duration 
$\sigma$ much shorter than the mechanical relaxation time (in the absence
of feedback), $\sigma \ll 1/\gamma_{m}$. 
In fact, one could use the following nonstationary strategy: prepare
at $t=0$ the mirror mode in the cooled stationary state of Section IV,
then suddenly turn off the feedback loop and perform the spectral 
measurement in the presence of the impulsive force
for a time $T_m $, such that $\sigma \ll T_m  \ll 1/\gamma_m$.
In such a way, the force spectrum is still
well reproduced, and the mechanical susceptibility is the one without feedback
(even though modified by the short measurement time $T_m \ll 1/\gamma_m$).
At the same time, the mechanical mode is far from equilibrium 
during the whole 
measurement, and its noise spectrum is different from the 
stationary form of Eq.~(\ref{qspedet}), being mostly determined by the 
{\em cooled} initial state. As long as $T_m \ll \gamma_m$, heating, that is,
the approach to the hotter equilibrium without feedback, will not
affect and increase too much the noise spectrum.
Therefore, one expects 
that as long as the measurement time is sufficiently short,
the SNR for the detection of the impulsive 
force (which has now to be evaluated using the most general expressions 
(\ref{signal2}) and (\ref{noise2})) can be significantly increased by 
this nonstationary strategy.

\begin{figure}[h]
\centerline{\epsfig{figure=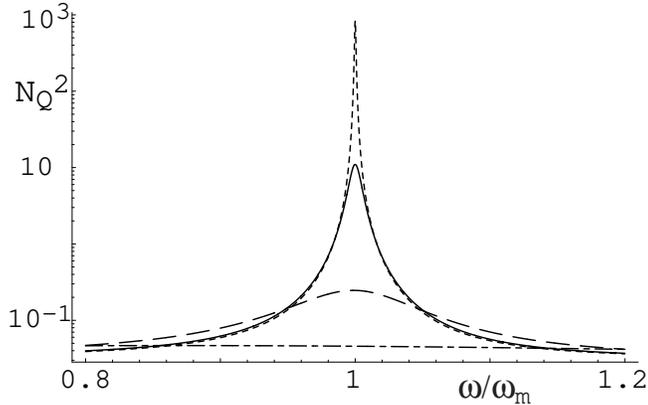,width=3.5in}}
\caption{\widetext 
Nonstationary noise spectrum $N_{Q}^{2}(\omega)=
N^{2}(\omega_{m})/4\eta \zeta \gamma_{m}T_{m}$ 
for different values of the measurement time,  
$\gamma_m T_m =10^{-1}$ (dotted line),
$\gamma_m T_m =10^{-2}$ (full line), $\gamma_m T_m =10^{-3}$ (dashed line),
$\gamma_m T_m =10^{-4}$ (dot-dashed line). 
The figure refers to the cold damping feedback scheme, but the curves 
are indistinguishable from that obtained with the stochastic cooling feedback,
using the same parameters, ${\cal Q}=10^4$, $\zeta =10$, $g_1=g_2=10^3$,
$k_BT/\hbar \omega_m = 10^5$, $\eta =0.8$.}
\label{nonstatm}
\end{figure}

\begin{figure}[h]
\centerline{\epsfig{figure=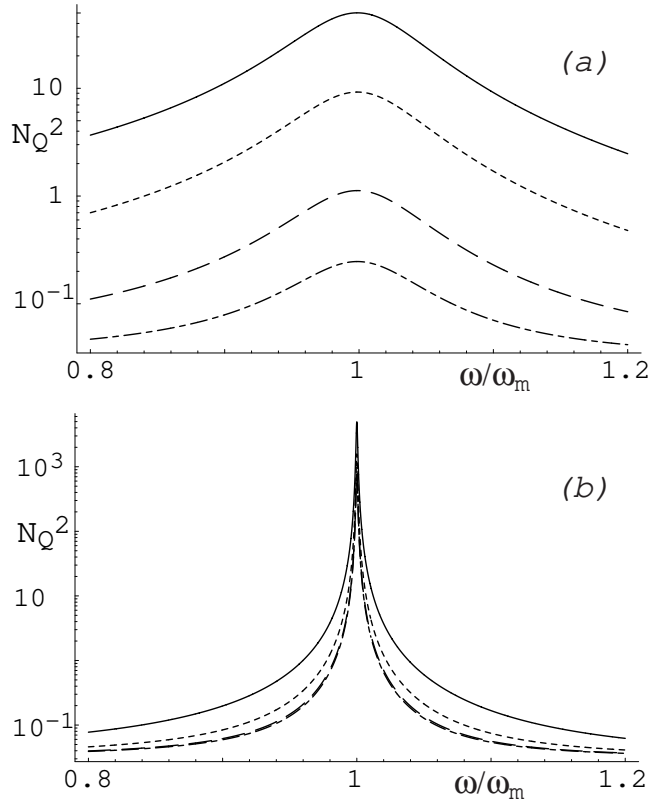,width=3.5in}}
\caption{\widetext 
Nonstationary noise spectrum 
$N_{Q}^{2}(\omega)$ 
for different values of the feedback gain,
$g_2=1$ (full line), $g_2=10$ (dotted line), $g_2=10^2$ (dashed),
$g_2=10^3$ (dot-dashed), with fixed measurement time,
$\gamma_m T_m=10^{-3}$ (a), and $\gamma_m T_m=10^{-1}$ (b). 
(a) corresponds to a strongly nonstationary
condition, in which the noise is significantly suppressed,
thanks to the cooled initial condition.
In (b) the stationary terms becomes important and the noise
reduction due to feedback cooling is less significant.
The figure refers to the cold damping feedback scheme, but the curves 
are indistinguishable from that obtained with the stochastic cooling feedback,
using the same parameters, ${\cal Q}=10^4$, $\zeta =10$,
$k_BT/\hbar \omega_m = 10^5$, $\eta =0.8$.}
\label{nonstag}
\end{figure}

It is instructive to evaluate explicitely the nonstationary
noise spectrum of Eq.~(\ref{noise2}) 
for the above measurement strategy.
Let us first consider the cold damping case, which gives more compact
expressions. Using Eq.~(\ref{QT}), one gets
\begin{equation}
C(t,t')=K(t)K(t') \langle Q^2\rangle_{st} +
\chi_0(t)\chi_0(t') \langle P^2\rangle_{st}+
\int_0^t dt_1\int_0^{t'} dt_2 \chi_0(t_1)\chi_0(t_2)c(t-t'-t_1+t_2),
\label{nonstacor}
\end{equation}
where $\chi_0(t)$ is the mechanical susceptibility in the absence of feedback
(see Eq.~(\ref{suscsc}) with $g_1=0$ or Eq.~(\ref{susccd}) with $g_2=0$),
$K(t)$ is given by Eq.~(\ref{K}) with $\chi_{cd}$ replaced by $\chi_0$,
$\langle Q^2\rangle_{st}$ and $\langle P^2\rangle_{st}$ are the stationary
values in the presence of feedback evaluated in Section IV, and $c(t)$ is the
cold damping noise correlation function introduced in Eqs.~(\ref{q21cd}) and
(\ref{corcd}).
This nonstationary correlation function has to be inserted in 
Eq.~(\ref{noise2}). Simple analytical results are obtained if we choose
the following filter function
\begin{equation}
F_{T_m}(t) =\theta(t) e^{-t/2T_m}
\label{filter}
\end{equation}
($\theta(t)$ is the Heavyside step function),
satisfying $\int dt F_{T_m}(t)^2 = T_m$. Using Eq.~(\ref{filter})
and rewriting $c(t)$ in terms of its Fourier transform ${\tilde c}(\omega)$,
one gets
\begin{eqnarray}
 N^{2}(\omega)&= &\frac{(8G\beta \eta)^{2}}{\gamma_{c}}\left[
  \left|{\tilde K}(\omega-i/2T_m)\right|^2 \langle Q^2\rangle_{st} +
\left|{\tilde \chi_0}(\omega-i/2T_m)\right|^2 \langle P^2\rangle_{st}
\right. \nonumber \\
&+& \left.
\left|{\tilde \chi_0}(\omega-i/2T_m)\right|^2
\int_{\infty}^{+\infty}\frac{d\omega'}{2 \pi}\frac{{\tilde c}(\omega')}{
\frac{1}{4T_m^2}+(\omega'-\omega)^2}\right] +\eta T_m .
  	\label{noisenosta}
  \end{eqnarray}
From Eq.~(\ref{K}), it is possible to see that ${\tilde K}(\omega)=
(i\omega +\gamma_m ){\tilde \chi}_0(\omega)/\omega_m$; then, using
Eq.~(\ref{corcd}) with $g_{cd}=0$, 
and the high temperature approximation $\coth(\hbar 
\omega /2k_B T) \simeq 2 k_B T/\hbar \omega$ for the Brownian noise, one
finally gets the following expression for nonstationary noise spectrum
for the cold damping feedback
\begin{equation}
 N^{2}(\omega)= \frac{(8G\beta \eta)^{2}}{\gamma_{c}}
 \left|{\tilde \chi_0}(\omega-i/2T_m)\right|^2 \left[
 \frac{\omega^2+(1/2T_m+\gamma_m)^2}{\omega_m^2}  \langle Q^2\rangle_{st} +
\langle P^2\rangle_{st}+ \gamma_{m}T_m \left( \frac{\zeta}{4}+
\frac{k_B T}{\hbar \omega_{m}}\right)\right] +\eta T_m .
  	\label{noisenosta2}
  \end{equation}
The corresponding noise spectrum for the stochastic cooling case 
can be obtained in a similar way. Using Eq.~(\ref{QTIME}), one gets
\begin{eqnarray}
C(t,t')&=& K_Q(t)K_Q(t') \langle Q^2\rangle_{st} +
\chi_0(t)\chi_0(t') \langle P^2\rangle_{st}+
\left[\chi_0(t)K_Q(t')+K_Q(t)\chi_0(t')\right] 
\frac{\langle QP+PQ\rangle_{st}}{2} \nonumber \\
&+& \int_0^t dt_1\int_0^{t'} dt_2 \chi_0(t_1)\chi_0(t_2)c(t-t'-t_1+t_2),
\label{nonstacorsc}
\end{eqnarray}
where $K_Q(t)$ is given by Eq.~(\ref{KQ}) 
(with $\chi_{sc}$ replaced by $\chi_0$),
$\langle Q^2\rangle_{st}$, $\langle P^2\rangle_{st}$ and 
$\langle QP+PQ\rangle_{st}$ are the stationary
values in the presence of stochastic cooling feedback evaluated 
in Section IV, and we have used the fact that, without
feedback, $c_1(t)=0$ and $c_2(t)=c(t) $ (see Eqs.~(\ref{corsc}) 
and (\ref{corcd})).
Inserting this nonstationary correlation function in 
Eq.~(\ref{noise2}), using Eq.~(\ref{filter}),
the fact that ${\tilde K}_Q(\omega)=
(i\omega +\gamma_m ){\tilde \chi}_0(\omega)/\omega_m$,
and again the high temperature approximation for the Brownian noise, one
finally gets
\begin{eqnarray}
 N^{2}(\omega)&=& \frac{(8G\beta \eta)^{2}}{\gamma_{c}}
 \left|{\tilde \chi_0}(\omega-i/2T_m)\right|^2 \left[
 \frac{\omega^2+(1/2T_m+\gamma_m)^2}{\omega_m^2}  \langle Q^2\rangle_{st} +
\langle P^2\rangle_{st}+ 
\frac{\gamma_m+1/2T_m}{\omega_m}\langle QP+PQ\rangle_{st} \right.\nonumber \\
 &+& \left.\gamma_{m}T_m \left( \frac{\zeta}{4}+
\frac{k_B T}{\hbar \omega_{m}}\right)\right] +\eta T_m .
  	\label{noisenostasc}
  \end{eqnarray}
Notice that the two noise spectra (\ref{noisenosta2}) and (\ref{noisenostasc})
are very similar, the only difference being in the initial stationary values,
whose explicit expression for the two feedback schemes is given in Section IV.
It is also easy to check that the stationary noise spectrum corresponding 
to the situation with no feedback is recovered in the limit of large 
$T_m$, as
expected, when the terms proportional to $\gamma_{m}T_m$ become dominant, and 
${\tilde \chi_0}(\omega-i/2T_m) \to {\tilde \chi_0}(\omega)$. In the opposite 
limit of small $T_m$ instead, 
the terms associated to the {\em cooled}, initial conditions 
are important, and since the terms proportional to $\gamma_{m}T_m$ are still
small, this means having a reduced, nonstationary noise spectrum.
This is clearly visible in Fig.~\ref{nonstatm}, where the nonstationary
noise spectrum, renormalized in order to have a position spectrum, 
$N_{Q}^{2}(\omega)=N^{2}(\omega)/4\eta \zeta \gamma_{m}T_{m}$, 
is plotted for different values of the
measurement time $T_m$, $\gamma_m T_m =10^{-1}$ (dotted line),
$\gamma_m T_m =10^{-2}$ (full line), $\gamma_m T_m =10^{-3}$ (dashed line),
$\gamma_m T_m =10^{-4}$ (dot-dashed line). The resonance peak is 
significantly suppressed for decreasing $T_m$, 
even if it is simultaneously widened, so that
one can even have a slight increase of noise out of resonance.
This figure is referred to the cold damping feedback scheme, but it is 
indistinguishable from that obtained with the stochastic cooling feedback,
using the same parameters (${\cal Q}=10^4$, $\zeta =10$, $g_1=g_2=10^3$,
$k_BT/\hbar \omega_m = 10^5$, $\eta =0.8$).
In fact, it can be checked that the two nonstationary 
noise spectra (\ref{noisenosta2}) and (\ref{noisenostasc}) 
differ significantly only at very low values of the mechanical quality factor
(${\cal Q} < 10^2$).
The effect of the terms depending upon the feedback-cooled initial 
conditions on the nonstationary noise is shown in 
Fig.~\ref{nonstag}, where the noise spectrum
is plotted for different values of the feedback gain at a fixed value
of $T_m$. In Fig.~\ref{nonstag}a, $N_{Q}^{2}(\omega)$ is plotted at 
$\gamma_m T_m=
10^{-3}$ for $g_2=1$ (full line), $g_2=10$ (dotted line), $g_2=10^2$ (dashed),
$g_2=10^3$ (dot-dashed). For this low value of $\gamma_m T_m$, 
the noise terms depending on the
initial conditions are dominant, and increasing the feedback gain implies
reducing the initial variances, and therefore an approximately 
uniform noise suppression at all frequencies.
In Fig.~\ref{nonstag}b, $N_{Q}^{2}(\omega)$ is instead plotted 
at $\gamma_m T_m=
10^{-1}$ for $g_2=1$ (full line), $g_2=10$ (dotted line), $g_2=10^2$ (dashed),
$g_2=10^3$ (dot-dashed). In this case, the feedback-gain-independent,
stationary terms become important,
and the effect of feedback on the noise spectrum
becomes negligible. Also in this case, Fig.~\ref{nonstag}
is valid for both stochastic cooling and cold damping schemes.

It is also possible to check from Eqs.~(\ref{noisenosta2}) and 
(\ref{noisenostasc}) that,
similarly to what happens for the stationary case, noise does not 
uniformly decrease for increasing feedback gain if the input 
power $\zeta$ is kept fixed, but there is an
optimal feedback gain, minimizing the noise at a given frequency and 
input power.

\begin{figure}[h]
\centerline{\epsfig{figure=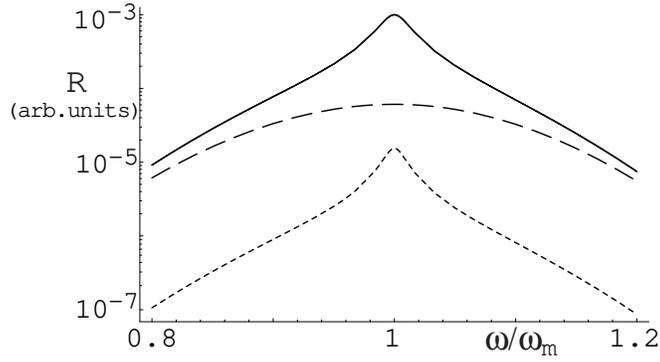,width=3.5in}}
\caption{\widetext 
Spectrum of the nonstationary SNR, ${\cal R}(\omega)$, with and without 
feedback cooling of the initial state. The full line refers to a 
nonstationary measurement, $\gamma_{m}T_{m}=10^{-3}$,
in the presence of feedback, $g=2\cdot 10^{3}$ (the two feedback 
schemes give indistinguishable curves); the dashed line refers to the 
no-feedback case, and with the same, short, measurement time 
$\gamma_{m}T_{m}=10^{-3}$. Finally, the dotted line refers to a ``standard 
measurement'', without feedback, and in the stationary limit
$\gamma_{m}T_{m}=10$. The other parameters are
$\omega_{f}=\omega_{m}$, $ \gamma_{m}\sigma = 10^{-4}$, 
$\gamma_{m}t_{1}=3 \cdot 10^{-4}$, 
${\cal Q}=10^{5}$, $\zeta=10$, $\eta =0.8$, $k_{B}T/\hbar 
\omega_{m}=10^{5}$.}
\label{snrnon1}
\end{figure}

\begin{figure}[h]
\centerline{\epsfig{figure=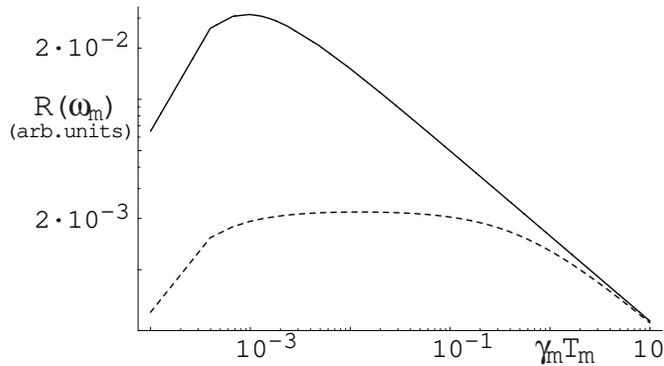,width=3.5in}}
\caption{\widetext 
Nonstationary SNR at resonance, ${\cal R}(\omega_{m})$, with and without 
feedback cooling of the initial state, plotted as a function of the 
rescaled measurement time $\gamma_{m}T_{m}$.
The full line refers to the case with feedback-cooled initial
conditions ($g=2\cdot 10^{3}$, the two feedback 
schemes give indistinguishable curves). The dotted line refers to the 
no-feedback case, $g=0$. The other parameters are the same as in 
Fig.~\protect\ref{snrnon1}.}
\label{snrnon2}
\end{figure}

The significant noise reduction attainable at short measurement times
$\gamma_{m}T_{m} \ll 1$ is not only due to the feedback-cooled 
initial conditions, but it is also caused by the effective 
reduction of the mechanical susceptibility given by the short 
measurement time, ${\tilde \chi}_{0}(\omega) 
\to {\tilde \chi}_{0}(\omega-i/2T_{m})$. 
This lowered susceptibility yields a simultaneous 
reduction of the signal at small measurement times $\gamma_{m}T_{m} 
\ll 1$, and therefore the behavior of the nonstationary SNR 
may be nontrivial. However, one expects that impulsive forces at least
can be satisfactorily 
detected using a short measurement time, because the noise can be kept 
very small and the corresponding sensitivity increased.
Let us check this fact considering the case of 
the impulsive force
\begin{equation}
f(t)=f_{0}\exp\left[-(t-t_{1})^{2}/2\sigma^{2}\right]
  	\cos\left(\omega_{f}t\right),
\label{force}
\end{equation}
where $\sigma$ is the force duration, $t_{1}$ its ``arrival time'', and
$\omega_{f}$ its carrier frequency. The corresponding SNR is obtained 
dividing the signal of Eq.~(\ref{signal}), evaluated with
Eq.~(\ref{filter}), by the nonstationary noise
spectra of Eqs.~(\ref{noisenosta2}) and (\ref{noisenostasc}), and it 
is shown in Figs.~\ref{snrnon1} and \ref{snrnon2}. As anticipated, the 
sensitivity of the optomechanical device is improved using
feedback in a nonstationary way. In Fig.~\ref{snrnon1}, the spectral SNR, 
${\cal R}(\omega)$, is plotted for different values of feedback gain 
and measurement time (as in the previous curves, the figures well 
describe both feedback schemes, because
they give indistinguishable results 
for ${\cal R}(\omega)$ in the physically relevant parameter region).
The full line refers to $g_1=g_{2}=g=2\cdot 10^{3}$ and 
$\gamma_{m}T_{m}=10^{-3}$, the dashed line to the situation with 
no feedback and the same measurement time, $g=0$ and 
$\gamma_{m}T_{m}=10^{-3}$; finally the dotted line refers to a 
``standard'' measurement, that is, no feedback and a stationary 
measurement, with a long measurement time, $\gamma_{m}T_{m}=10$. 
The proposed nonstationary measurement scheme,
``cool and measure'', gives the highest sensitivity. This is 
confirmed also by Fig.~\ref{snrnon2}, where the SNR at resonance,
${\cal R}(\omega_{m})$, when feedback cooling is used with
$g=2\cdot 10^{3}$ (full line), and without feedback cooling
(dotted line), is plotted as a function of the rescaled
measurement time $\gamma_{m}T_{m}$. The preparation of the mirror in 
the cooled initial state yields a better sensitivity for any 
measurement time. As expected, the SNR in the presence of 
feedback approaches that without feedback in the stationary limit
$\gamma_{m}T_{m} \gg 1$, when the effect of the initial cooling 
becomes irrelevant.
Both Fig.~\ref{snrnon1} and \ref{snrnon2} refer 
to a resonant ($\omega_{f}=\omega_{m}$)
impulsive force with $ \gamma_{m}\sigma = 10^{-4}$ and 
$\gamma_{m}t_{1}=3 \cdot 10^{-4}$, while the other parameters are
${\cal Q}=10^{5}$, $\zeta=10$, $\eta =0.8$, $k_{B}T/\hbar 
\omega_{m}=10^{5}$.

The proposed nonstationary strategy 
can be straightforwardly applied whenever the ``arrival 
time'' $t_{1}$ of the impulsive force
is known: feedback has to be 
turned off just before the arrival of the force. However, the 
scheme can be easily adapted also to the case of an impulsive force with an 
{\em unknown arrival time}, as for example, that of
a gravitational wave passing through an interferometer. In this case 
it is convenient to repeat the process many times, i.e., 
subject the oscillator to cooling-heating cycles. 
Feedback is turned off for a time $T_{m}$ during which the spectral measurement 
is performed and the oscillator starts heating up. Then feedback is 
turned on and the oscillator is cooled, and then the process is 
iterated. This cyclic cooling strategy improves the
sensitivity of gravitational wave detection 
provided that the cooling time $T_{cool}$,
which is of the order of $1/\left[\gamma_{m}(1+g_{i})\right]$, is much 
smaller than $T_{m}$, which is verified at sufficiently large gains.
Cyclic cooling has been proposed, in a qualitative way, 
to cool the violin modes of 
a gravitational waves interferometer in \cite{PINARD}, and its 
capability of improving the high-sensitive detection of impulsive forces
has been first shown in \cite{LETTER}. 
In the case of a random, uniformly distributed, arrival time
$t_1$ and in the impulsive limit $\sigma \ll T_{m}$, 
the performance of the cyclic cooling scheme is well characterized by 
a time averaged SNR, i.e., 
\begin{equation}
\langle {\cal R}(\omega)\rangle = 
\frac{1}{T_{m}+T_{cool}}\left\{\int_{0}^{T_{m}}dt_{1}
{\cal R}(\omega,t_{1})+\int_{T_{m}}^{T_{m}+T_{cool}}dt_{1}
{\cal R}(\omega,t_{1})_{cool}\right\},
	\label{snrmedio}
\end{equation}
where ${\cal R}(\omega,t_{1})$ is the nonstationary SNR at a given 
force arrival time $t_{1}$ discussed in this section, and ${\cal 
R}(\omega,t_{1})_{cool}$ is the nonstationary SNR one has during the 
cooling cycle, which means with feedback turned on and with uncooled 
initial conditions. It is easy to understand that ${\cal 
R}(\omega,t_{1})_{cool} \ll {\cal 
R}(\omega,t_{1})$, and, since it is also $T_{cool} \ll T_{m}$, 
the second term in Eq.~{\ref{snrmedio}) can be neglected, so that
\cite{LETTER},
\begin{equation}
\langle {\cal R}(\omega)\rangle \simeq 
\frac{1}{T_{m}+T_{cool}}\int_{0}^{T_{m}}dt_{1}
{\cal R}(\omega,t_{1}).
	\label{snrmedio2}
\end{equation}
 
This time-averaged SNR can be significantly improved 
by cyclic cooling, as it is shown in Fig.~\ref{snrme}, 
where $\langle {\cal R}(\omega)\rangle$ is plotted both with and 
without feedback. The full line describes 
the time-averaged SNR subject to cyclic feedback-cooling 
with $g=2\cdot 10^{3}$, $\gamma_{m}T_{m}=10^{-3}$, and
$T_{cool}=10^{-3}T_{m}$. In the absence of feedback, in the case of 
an impulsive 
force with unknown arrival time and duration $\sigma$, 
the best strategy is to perform repeated measurements of 
duration $T_{m}$ without any cooling stage. The measurement time 
$T_{m}$ can be optimized considering that it has to be longer
than $\sigma$, and at the same time it has not to be too long, in 
order to have a good SNR (see the dotted line in Fig.~\ref{snrnon2}).
In this case, the time-averaged SNR can be written as
\begin{equation}
\langle {\cal R}_{0}(\omega)\rangle \simeq 
\frac{1}{T_{m}}\int_{0}^{T_{m}}dt_{1}
{\cal R}_{0}(\omega,t_{1}),
	\label{snrmedio3}
\end{equation}
where ${\cal R}_{0}(\omega,t_{1})$ is the SNR evaluated for $g=0$.
The dashed line in Fig.~\ref{snrme} refers to this case without 
feedback, and with $\gamma_{m}T_{m}=10^{-3}$. 
The other parameter values
are the same as in Figs.~\ref{snrnon1} and \ref{snrnon2} and in this 
case, cyclic cooling provides
an improvement at resonance by a factor $16$ with respect to 
the case with no feedback.
As suggested in Ref.~\cite{PINARD},
one could use nonstationary cyclic feedback to cool the violin modes 
in gravitational-wave interferometers, which have sharp resonances 
within the detection band. One expects that single gravitational
bursts, having a duration smaller than the cooling cycle period, 
could be detected in this way.

\begin{figure}[h]
\centerline{\epsfig{figure=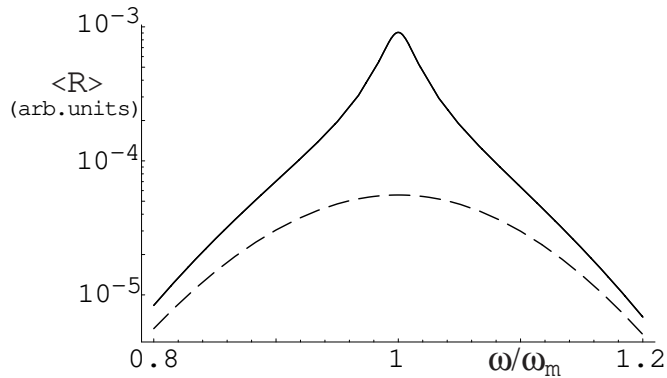,width=3.5in}}
\caption{Time averaged spectral SNR 
with and without cyclic cooling. The full line refers to 
cyclic cooling with $\gamma_{m}T_{m}=10^{-3}$,
$g=2\cdot 10^{3}$, and $T_{cool}=10^{-3}T_{m}$ (the two feedback 
schemes give indistinguishable curves). The dashed line refers to the 
no-feedback case, with the same measurement time 
$\gamma_{m}T_{m}=10^{-3}$ (see Eq.~(\protect\ref{snrmedio3})).
The other parameters are
$\omega_{f}=\omega_{m}$, $ \gamma_{m}\sigma = 10^{-4}$,  
${\cal Q}=10^{5}$, $\zeta=10$, $\eta =0.8$, $k_{B}T/\hbar 
\omega_{m}=10^{5}$.}
\label{snrme}
\end{figure}

\section{Conclusions}

We have studied how quantum feedback schemes can be used to reduce 
thermal noise and improve the sensitivity of optomechanical devices.
We have analysed in detail the stochastic cooling scheme introduced in
Ref.~\cite{MVTPRL} and the cold damping scheme experimentally implemented
in Ref.~\cite{HEIPRL,PINARD}. We have seen that the two schemes are 
physically analogous, even though they show some differences. In both 
cases, the main effect of feedback is the increase of mechanical damping, 
accompanied by the introduction of a controllable, measurement-induced, 
noise. The increase of damping means reduction of the susceptibility 
at resonance, and the consequent suppression of the 
resonance peak in the noise spectrum. Stochastic cooling feedback 
differs form cold damping in the fact that it has the supplementary 
effect of increasing the mechanical frequency. This means that, while
cold damping achieves thermal noise reduction only around resonance, 
stochastic cooling is able to reduce noise even at very low frequencies, 
out of resonance. We have also shown that both schemes are able 
to achieve the ultimate quantum limit of ground state cooling
(see also \cite{COURTY} for the cold damping case).
For both feedback schemes, ground state cooling
is reached in the limit of very large 
feedback gain, ideal homodyne detection, and very large input power. 
In the stochastic cooling case, however, also the additional condition 
of very large mechanical quality factor is needed (see also 
\cite{RON}), so that cooling is much more easily achieved in the cold 
damping case. In the limit of very large gain and input power, but 
with fixed mechanical quality factor, stochastic cooling feedback 
is instead able to achieve steady state position squeezing, that is,
one can beat the standard quantum limit $\langle Q^{2}_{st} \rangle < 
1/4$. Finally stochastic cooling is also able to produce stationary 
contractive states \cite{YUEN}. Reaching these 
quantum limits in optomechanical sytems 
is experimentally very difficult but it would be extremely important, 
because it would be a genuine manifestation of quantum 
mechanics for a macroscopic mechanical degree of freedom. 

We have also analysed the sensitivity of the optomechanical device 
in the case of position spectral measurements for the detection of 
weak forces. Even though both feedback schemes are not able to 
improve the sensitivity of stationary measurements, we 
have shown how feedback can be used in a nonstationary way in order to 
increase of the SNR in the case of impulsive forces. If the 
arrival time of the classical force is known, one has to keep the 
mirror mode cooled by feedback, and then turn off the feedback just 
before the arrival of the force. The mirror therefore responds to the 
force with its intrinsic susceptibility, not suppressed by the 
feedback, and with a nonstationary noise, reduced by the feedback.
The SNR is increased as long as the measurement time $T_{m}$ is 
longer than the force duration $\sigma$, but much smaller than the mechanical 
relaxation time, that is, $\sigma \ll T_{m} \ll 1/\gamma_{m}$.
This nonstationary strategy can be well adapted to the case
of a force with an unknown arrival time, as for example, gravitational 
waves. In this case, the cooling and measurement steps has to be
cyclically repeated, and the performance of cyclic cooling can be 
characterized by a SNR averaged over the force arrival time. 
This time-averaged SNR can be significantly improved by cyclic 
cooling, thanks also to the fact that the cooling time can be
made very small using very large feedback gains $g$,
because it is $T_{cool} \simeq \left[\gamma_{m}(1+g)\right]^{-1}$.
Differently from ground state cooling, the experimental
implementation of these nonstationary strategy is feasible with 
current technology, and it may be 
useful not only for optomechanical devices, but also for 
microelectromechanical systems.

\end{document}